\UseRawInputEncoding
\documentclass[epj]{svjour}
\usepackage{graphics}
\usepackage{amsmath}
\usepackage{amsfonts}
\usepackage{amssymb}
\usepackage{xcolor}
\usepackage[normalem]{ulem}
\begin{document}
\title{Rotating black holes in de Rham-Gabadadze-Tolley massive gravity: Analytic calculation procedure}

\author{Ping Li\inst{1,2} \and Jiang-he Yang\inst{1,3}
}                     
\offprints{}          
\institute{College of Mathematics and Physics, Hunan University of Arts and Sciences, 3150 Dongting Dadao, Changde City, Hunan Province 415000, China \and 
Hunan Province Key Laboratory Integration and Optical Manufacturing Technology, 3150 Dongting Dadao, Changde City, Hunan Province 415000, China  \and
Center for Astrophysics, Guangzhou University, 230 West Ring Road, Guangzhou, Guangdong Province 510006, China
}
\date{Received: date / Revised version: date}
%
\abstract{
In this paper, we explore the solutions of rotating black holes within the framework of de Rham-Gabadadze-Tolley (dRGT) massive gravity. We provide a detailed, step-by-step analytical derivation of these solutions. Our solutions are characterized by several parameters: mass $M$ , electric charge $Q_{*}$, angular momentum $a$, and a graviton mass $m$. This graviton mass term incorporates both a cosmological constant $\Lambda$ and a St\"uckelberg charge $S_{*}$ into the black hole parameters. These solutions may serve as potential candidates for astrophysical black holes.
\PACS{
      {04.20.Jb}{discribing text of that key}   \and
      {04.50.Kd}{discribing text of that key} \and
      {04.70.Bw}{discribing text of that key} \and
      {02.30.Jr}{discribing text of that key}
     } 
} 
\maketitle
\section{Introduction}
The introduction of a mass term for gravity presents a formidable challenge. The linear mass term is plagued by the van Dam-Veltman-Zakharov (vDVZ) discontinuity \cite{VanDam1970}. On the other hand, the nonlinear mass term may propagate the sixth mode on a curved background, known as the Boulware-Deser (BD) ghost \cite{Boulware1972}. To circumvent these issues, Comelli and his cooperators try to introduce mass terms without Lorentz symmetry \cite{Comelli2012,Comelli2013}. We presented a detailed study of the spherically symmetric solutions in Lorentz breaking massive gravity \cite{Li2016-2}. Concurrently, de Rham, Gabadadze, and Tolley proposed a healthy Lorentz invariant massive gravity model \cite{deRham2011} in 2011.

In dRGT theory, the leading contribution to the decoupling limit is a total derivative which prevents the emergence of ghosts. Beyond the decoupling limit, the lapse function is a Lagrange multiplier that constraint the sixth mode in an arbitrary background. Thus, the dRGT theory described a massive spin-2 particles that propagate 5 healthy degrees of freedom. A comprehensive review of massive gravity is available in \cite{deRham2014}. If graviton possesses mass, the gravitational waves would travel at a speed different from light. The recent observation of GW150914 by LIGO \cite{Baker2017} has placed an upper bound on the graviton mass $m_g\leq1.2\times10^{-22} eV$. For more about the graviton mass bounds, you may see Ref. \cite{deRham2017}. Fortunately, the dRGT theory can produce a small graviton mass technically natural \cite{deRham2013}.

Numerous studies have explored black hole solutions within the dRGT theory. Nieuwenhuizen obtained the  Sch-warzschild-de Sitter and Reissner-Nordstr\"{o}m-de Sitter black holes \cite{Nieuwenhuizen2011}. Cai \textit{et. al.} also derived the static charged black hole solutions \cite{Cai2013}. De Rham and his colleagues \cite{Berezhiani2011} highlighted the existence of a new basic invariant $I^{ab}=g^{\alpha\beta}\partial_{\alpha}\phi^a\partial_{\beta}\phi^b$, whose singularities could pose a problem for fluctuations. In the unitary gauge, it is evident that $I^{ab}=g^{\alpha\beta}\delta_{\alpha}^a\delta_{\beta}^b$ would exhibit a singularity if $g_{\alpha\beta}$ contains an event horizon. Consequently, there are no self-consistent static spherically symmetric (SSS) solutions in the unitary gauge. In our previous work \cite{Li2016-1}, we utilized the nonunitary gauge to search for SSS solutions and discovered seven solutions, including black hole with the St\"{u}ckelberg hair. We also derive the Vaidya solution in Ref. \cite{Li2016-3} and investigated the superradiant instabilities for charged furry black holes in dRGT theory \cite{Li2020}. Many papers have also studied the dRGT theory with degenerate reference metrics \cite{Cai2015,Adams2015,Xu2015,Cao2016,zhang2016,zhang2017}. And Ref. \cite{Jafari2017} calculate the SSS solutions in arbitrary dimensions.

Recently, we extended Chandrasekhar's method of calculating rotating black holes into $f(R)$ theory \cite{Li2022,Li2024-1,Li2024-2}. In this paper, we further extend this analytical method into the dRGT massive gravity. We proved that arbitrary nonunitary gauge $\phi^a$ with the Minkowski reference metric $\eta_{ab}$ equal to the unitary gauge $\bar{\phi}^a=x^{\alpha}\delta^a_{\alpha}$ with a curved reference metric $\bar{f}_{ab}$. Thus, instead of chosen the ans\"{a}tz of the St\"{u}ckelberg
field $\phi^a$, we directly choose the axisymmetric reference metric $f_{ab}$ with a unitary gauge $\phi^a=x^{\alpha}\delta^a_{\alpha}$. The field equations with the electromagnetic fields can be simplified to the Ernst equations. The shortest and simplest route to obtain rotating solutions is to use the conjugate variables. We solve the homogeneous equations and the non-homogeneous equations with the conjugate variables respectively. And the solution that beyond the Kerr-Newman family is obtained.

This paper is organized as follows: Sec. II review the field equation of dRGT theory \cite{deRham2011}. In Sec. III, we give a general stationary axisymmetric metric and calculate the Einstein tensor $G_{ab}$. In Sec. IV, we discussed the electromagnetic field equations in the stationary axisymmetric background and the energy-momentum of electromagnetic fields. In Sec. V, we proved that arbitrary nonunitary gauge $\phi^a$ with the Minkowski reference metric $\eta_{ab}$ equal to the unitary gauge $\bar{\phi}^a=x^{\alpha}\delta^a_{\alpha}$ with a curved reference metric $\bar{f}_{ab}$. And we derive the matrix square root $\gamma$ by using the Cayley-Hamilton theorem and calculate the $T^{(\mathcal{K})}_{ab}$. In Sec. VI, we simplify the metric to a highly symmetric form by choosing the gauge freedom. In Sec. VII, we derive the Ernst equations. In Sec. VIII, we transform the variables into a conjugate metric. In Sec. IX, we solve the homogeneous equations of the Ernst equations. In Sec. X, we obtain the non-homogeneous solutions. The last section is a detailed conclusion and discussion.

In this paper, the Greek letters $\alpha,\beta,\sigma,...=0,1,2,3$ represent indices of the natural coordinate base, and the Latin letters $a,b,c,d=0,1,2,3$ represent indices of moving frame. The Greek letters are lowered or upped by the metric $g_{\alpha\beta}$, and the Latin letters are lowered or upped by the metric $\eta_{ab}=\text{diag}(-1,1,1,1)$.

\section{The modified Einstein equations in dRGT theory}
The action of the de Rham-Gabadadze-Tolley theory is given by
\begin{equation}\label{action}
S=\int d^4x\sqrt{-g}(R+m^2U(g,\phi^a))+S_{m},
\end{equation}
where $R$ is the Ricci scalar, and $U$ is a potential if the graviton have mass $m$. $S_m$ is the action of matter and $G=1$. The potential $U$ in four-dimensional spacetime is composed of three parts,
\begin{equation}\label{potential}
 U(g,\phi^a)=U_2+\alpha_3U_3+\alpha_4U_4,
\end{equation}
where
\begin{align}
  U_2&= [\mathcal{K}]^2-[\mathcal{K}^{2}],  \\
  U_3& =[\mathcal{K}]^3-3[\mathcal{K}][\mathcal{K}^{2}]+2[\mathcal{K}^{3}],\\
  U_4&=[\mathcal{K}]^4-6[\mathcal{K}]^2[\mathcal{K}^{2}]+8[\mathcal{K}][\mathcal{K}^{3}]+3[\mathcal{K}^{2}]^2-6[\mathcal{K}^{4}],\\
  \mathcal{K}^{\alpha}{}_{\beta}&=\delta^{\alpha}_{\beta}-\gamma^{\alpha}{}_{\beta},
\end{align}
with $\alpha_3,\alpha_4$ being dimensionless parameters. Here the square brackets $[A]$ denote the trace of matrix $A_{\alpha\beta}$. The abbreviation of $\gamma^{i}$ stands for the contraction of $i$ copies of $\gamma$,
\begin{equation}
  \gamma^{i}{}^{\alpha}{}_{\beta}=\gamma^{\alpha}{}_{\alpha_1}\gamma^{\alpha_1}{}_{\alpha_2}...\gamma^{\alpha_{i-1}}{}_{\beta}.
\end{equation}
We also set
\begin{equation}
  \gamma^{0}{}^{\alpha}{}_{\beta}=\delta^{\alpha}_{\beta}.
\end{equation}
And the trace of $\gamma^{i}{}^{\alpha}{}_{\beta}$ is defined as $[\gamma^{i}]=g^{\alpha\beta}\gamma^{i}_{\alpha\beta}$. In the original Ref. \cite{deRham2011}, the matrix $\gamma^{2}$ is defined by the expression
\begin{equation}\label{gamma}
  \gamma^{2}{}^{\alpha}{}_ {\beta}=g^{\alpha\sigma}\partial_{\sigma}\phi^a\partial_{\beta}\phi^bf_{ab},
\end{equation}
where $\phi^a$ are the St\"{u}ckelberg fields. $\gamma^{\alpha}{}_{\beta}$ is the matrix square root of $\gamma^{2}{}^{\alpha}{}_ {\beta}$. The non-degenerate matrix $f_{ab}$ is named the reference (or fiducial) metric. Correspondingly, we also call $g_{\alpha\beta}$ the dynamic metric.

Variation of the action (\ref{action}) with respect to metric $g_{\alpha\beta}$ leads to the modified Einstein equations
\begin{equation}\label{ModEin}
 G_{\alpha\beta}+m^2T^{(\mathcal{K})}_{\alpha\beta}=8\pi T^{(m)}_{\alpha\beta},
\end{equation}
where
\begin{equation}
  T^{(\mathcal{K})}_{\alpha\beta}=\frac{1}{\sqrt{-g}}\frac{\delta(\sqrt{-g}U)}{\delta g^{\alpha\beta}},
\end{equation}
and $T^{(m)}_{\alpha\beta}$ represents the energy-momentum tensor of mater. It should be pointed out that the following calculations are acually using the tetrad components $a$, not $\alpha$. Due to the Bianchi identity $\nabla_{\alpha}G^{\alpha\beta}=0$ and the conservation of energy-momentum tensor $\nabla_{\alpha}T^{(m)}{}^{\alpha\beta}=0$, there is $\nabla_{\alpha}T^{(\mathcal{K})}{}^{\alpha\beta}=0$. As is pointed in Ref.\cite{Cao2016}, the variation of $[\gamma^{i}]$ is given by the relationship
\begin{equation}
  \delta [\gamma^{i}]=\frac{i}{2}\gamma^{i}_{\alpha\beta}\delta g^{\alpha\beta}.
\end{equation}
Then, one can easily derive
\begin{align}
 T^{(\mathcal{K})}{}^{\alpha}{}_{\beta}&=(-\frac{U}{2}+3+6(\alpha_3+\alpha_4))\delta^{\alpha}_{\beta}+([\gamma]-3)\gamma^{\alpha}{}_{\beta}-\gamma^{2}{}^{\alpha}{}_{\beta}\nonumber\\
 &-\frac{3\alpha_3}{2}(
 (6-4[\gamma]+\mathcal{U}_2)\gamma^{\alpha}{}_{\beta}
 -2([\gamma]-2)\gamma^{2}{}^{\alpha}{}_{\beta} +2\gamma^{3}{}^{\alpha}{}_{\beta})\nonumber\\
 &-2\alpha_4((6-6[\gamma]+3\mathcal{U}_2-\mathcal{U}_3)\gamma^{\alpha}{}_{\beta}+3(2-2 [\gamma]+\mathcal{U}_2)\gamma^{2}{}^{\alpha}{}_{\beta}\nonumber\\
 &+6(1-[\gamma])\gamma^{3}{}^{\alpha}{}_{\beta}+6\gamma^{4}{}^{\alpha}{}_{\beta} ),\label{TK}
\end{align}
where
\begin{align}
  \mathcal{U}_2 & = [\gamma]^2-[\gamma^{2}],\\
  \mathcal{U}_3 & =[\gamma]^3-3[\gamma][\gamma^{2}] +2[\gamma^{3}],\\
  \mathcal{U}_4&=[\gamma]^4-6[\gamma^{2}][\gamma]^2+8[\gamma][\gamma^{3}] +3[\gamma^{2}]^2-6[\gamma^{4}].
\end{align}
In the actual calculation process, the following equations are also important
\begin{equation}
  \mathcal{K}^{n}{}^{\alpha}{}_{\beta}=\sum_{k=0}^{n}(-1)^k\left(
                                                                           \begin{array}{c}
                                                                             n \\
                                                                             k \\
                                                                           \end{array}
                                                                         \right)
                                                                         \gamma^{k}{}^{\alpha}{}_ {\beta},
\end{equation}
Therefore, the traces of $\mathcal{K}^{i}{}^{\alpha}{}_{\beta}$ are given by
\begin{align}
  [\mathcal{K}] & =4-[\gamma], \\
  [\mathcal{K}^{2}] & =4-2[\gamma]+[\gamma^{2}],\\
  [\mathcal{K}^{3}] & =4-3[\gamma]+3[\gamma^{2}]-[\gamma^{3}],\\
  [\mathcal{K}^{4}] & =4-4[\gamma]+6[\gamma^{2}]-4[\gamma^{3}]+[\gamma^{4}].
\end{align}
Insert these equations into Eqs. (\ref{potential}), we have
\begin{align}
 U_2 & =12-6[\gamma]+ \mathcal{U}_2,\\
 U_3 & =24-18[\gamma]+6\mathcal{U}_2-\mathcal{U}_3,\\
 U_4&=24-24[\gamma]+12 \mathcal{U}_2 -4\mathcal{U}_3+\mathcal{U}_4.
\end{align}

\section{Stationary axisymmetric spacetime}
To derive the rotating black hole solutions, we consider the dynamic metric $g_{\alpha\beta}$ to process a stationary axial symmetry. It is convenient to take as two of coordinates the time $x^0=t$ and the azimuthal angle $x^1=\varphi$ about the axis of symmetry. At that time, we leave $(x^2,x^3)$ to be the remaining spatial coordinates. The stationary and the axisymmetric character of the spacetime require that the metric coeffiecients be independent of $t$ and $\varphi$, i.e.
\begin{equation}
  g_{\mu\nu}=g_{\mu\nu}(x^2,x^3).
\end{equation}
Besides, we also requrie that the spacetime is ivariant under simultaneous inversion of $t\rightarrow -t$ and $\varphi\rightarrow-\varphi$. The physical reason is that the energy-momumentum tensor of gravitational source would have the stated invariance provided that the motions of source are purely rotational about the axis of symmetry. The additional invariance requires that
\begin{equation}
 g_{02}=g_{03}=g_{12}=g_{13}=0.
\end{equation}
Then, under the assumptions made, the metric have the form
\begin{align}\label{AxiForm}
  ds^2&=g_{00}dt^2+2g_{01}dtd\varphi+g_{11}d\varphi^2\nonumber\\
  &+g_{22}(dx^2)^2+2g_{23}dx^2dx^3+g_{33}(dx^3)^2.
\end{align}
Notice that the two-dimensional space $(x^2,x^3)$ can always be brought to the diagonal form
\begin{equation}
 \pm e^{2\mu}[(dx^2)^2+(dx^3)^2],
\end{equation}
which is shown in Ref. \cite{Chandrasekhar:1983}. At last, we shall write the stationary axisymmetric metric in the form 
\begin{align}
   ds^2=&-e^{2\nu(x^2,x^3)}dt^2+e^{2\psi(x^2,x^3)}(d\varphi-q(x^2,x^3)dt)^2\nonumber\\
   &+e^{2\mu_2(x^2,x^3)}(dx^2)^2
  +e^{2\mu_3(x^2,x^3)}(dx^3)^2,  \label{dym}
\end{align}
where all unknown variables $\nu,\psi,\mu_2,\mu_3,q$ are the functions of $(x^2,x^3)$. In the above metric (\ref{dym}), we have left the functions $\mu_2$ and $\mu_3$ to be different. At a later stage, we may avail ourselves of this gauge-freedom to restrict $\mu_2$ and $\mu_3$ by a coordinate condition which may appear advantageous.

Using the metric (\ref{dym}), one can obtain the Einstein tensor $G_{\alpha\beta}$. To perform a basic calculation, we can apply the structural equation of a manifold. The Riemannian manifold are fully described by the frame 1-forms $\omega^a$ and the Riemann Levi-Civita (RLC) connection forms $\omega_{ab}$. We consider the spacetime is a torsion-free and metric compatible Riemannian manifold. Thus, the Cartan's equation of structure are given by
\begin{align}
  \mathrm{d} \omega^{a}+\omega^{a}{}_{b}\wedge \omega^{b} & =0,\label{CartanS1} \\
 \mathrm{d}\omega^{a}{}_{b}+ \omega^{a}{}_{c}\wedge \omega^{c}{}_{b}& =\Omega^{a}{}_{b},\label{CartanS2}
\end{align}
where $\mathrm{d}$ is the exterior differential operator and $\wedge$ stands for Cartan wedge. $\Omega^{a}{}_{b}$ are curvature 2-forms. After direct and tedious mathematical calculation \cite{Li2022,Li2024-1,Li2024-2}, one obtain nonzero components of $G_{ab}$ as follows:
\begin{align}
 G^{0}{}_{0}  =&e^{-2\mu_2}\left(\psi_{,2,2}+\psi^2_{,2}+\psi_{,2}(\mu_3-\mu_2)_{,2}\right)
 +e^{-2\mu_3}\big(\psi_{,3,3}\nonumber\\&+\psi^2_{,3}+\psi_{,3}(\mu_2-\mu_3)_{,3}\big)
+e^{-\mu_2-\mu_3}\big((e^{\mu_2-\mu_3}\mu_{2,3})_{,3}\nonumber\\&+(e^{\mu_3-\mu_2}\mu_{3,2})_{,2}
\big) +\frac{1}{4}e^{2(\psi-\nu-\mu_2)}q_{,2}^2+\frac{1}{4}e^{2(\psi-\nu-\mu_3)}q_{,3}^2, \label{G00}\\
 G^{1}{}_{1} = &e^{-2\mu_2}\left(\nu_{,2,2}+\nu^2_{,2}+\nu_{,2}(\mu_3-\mu_2)_{,2}\right)
 +e^{-2\mu_3}\big(\nu_{,3,3}\nonumber\\
 &+\nu^2_{,3}+\nu_{,3}(\mu_2-\mu_3)_{,3}\big)+e^{-\mu_2-\mu_3}\big((e^{\mu_2-\mu_3}\mu_{2,3})_{,3}\nonumber\\
 &+(e^{\mu_3-\mu_2}\mu_{3,2})_{,2}
\big) -\frac{3}{4}e^{2(\psi-\nu-\mu_2)}q_{,2}^2-\frac{3}{4}e^{2(\psi-\nu-\mu_3)}q_{,3}^2, \label{G11}\\
 G^{2}{}_{2}=& e^{-2\mu_2}\left(\psi_{,2}\nu_{,2}+(\psi+\nu)_{,2}\mu_{3,2}\right)
 +e^{-2\mu_3}\big((\psi+\nu)_{,3,3}\nonumber\\
 &+\nu^2_{,3}+\psi^2_{,3}+\psi_{,3}\nu_{,3}-(\psi+\nu)_{,3}\mu_{3,3}\big)\nonumber\\
 & +\frac{1}{4}e^{2(\psi-\nu-\mu_2)}q_{,2}^2-\frac{1}{4}e^{2(\psi-\nu-\mu_3)}q_{,3}^2, \label{G22}\\
  G^{3}{}_{3}=& e^{-2\mu_3}\left(\psi_{,3}\nu_{,3}+(\psi+\nu)_{,3}\mu_{2,3}\right)
+e^{-2\mu_2}\big((\psi+\nu)_{,2,2}\nonumber\\
 &+\nu^2_{,2}+\psi^2_{,2}+\psi_{,2}\nu_{,2}-(\psi+\nu)_{,2}\mu_{2,2}\big)
 -\frac{1}{4}e^{2(\psi-\nu-\mu_2)}q_{,2}^2\nonumber\\
 &+\frac{1}{4}e^{2(\psi-\nu-\mu_3)}q_{,3}^2, \label{G33}\\
 G^{0}{}_{1}=&-\frac{1}{2}e^{-(2\psi+\mu_2+\mu_3)}\big((e^{3\psi-\nu+\mu_3-\mu_2}q_{,2})_{,2}\nonumber\\
 &
 + (e^{3\psi-\nu+\mu_2-\mu_3}q_{,3})_{,3}\big),\label{G01}\\
 G^{2}{}_{3}=&-e^{-\mu_2-\mu_3}\big((\psi+\nu)_{,2,3}+\nu_{,2}\nu_{,3}+\psi_{,2}\psi_{,3}-(\psi+\nu)_{,2}\mu_{2,3}\nonumber\\&-(\psi+\nu)_{,3}\mu_{3,2}\big)
 +\frac{1}{2}e^{2(\psi-\nu)-\mu_2-\mu_3}q_{,2}q_{,3}.\label{G23}
\end{align}

Notice that we actually obtained the tetrad components $G_{ab}$ of the Einstein tensor. The frame 1-forms corresponding to metric (\ref{dym}) are $\omega^a=e^a{}_{\alpha}dx^{\alpha}$, where
\begin{equation}
 e^a{}_{\alpha}=\left(
                  \begin{array}{cccc}
                   e^{\nu} & 0 & 0 & 0 \\
                    -qe^{\psi} & e^{\psi} & 0 & 0 \\
                    0 & 0 & e^{\mu_2} & 0 \\
                    0 & 0 & 0 & e^{\mu_3} \\
                  \end{array}
                \right).
\end{equation}
Given any tensor $T^{\alpha...}{}_{\beta...}$, the tetrad components $T^{a...}{}_{b...}$ can be obtained by the projection
\begin{equation}
  T^{a...}{}_{b...}=T^{\alpha...}{}_{\beta...}e_{\alpha}{}^{a}\cdots e^{\beta}{}_{b}\cdots.
\end{equation}

\section{The electromagnetic field in stationary axisymmetric space-time}
In this section, we consider the electromagnetic field in stationary axisymmetric background. Given the action for electromagnetic fields $S_{m}=-\frac{1}{16\pi}\int \sqrt{-g}F_{\alpha\beta}F^{\alpha\beta}d^4x$, the energy-momentum is expressed as
\begin{equation}\label{EMT}
  T^{(m)}_{\alpha\beta}=\frac{1}{4\pi}\bigg(F_{\alpha\sigma}F_{\beta}{}^{\sigma}-\frac{1}{4}g_{\alpha\beta}F_{\sigma \kappa}F^{\sigma\kappa}\bigg),
\end{equation}
where $F_{\alpha\beta}=\partial_{\alpha}A_{\beta}-\partial_{\beta}A_{\alpha}$, and $A_{\alpha}$ is the vector potential of electromagnetic fields. In the stationary axisymmetric spacetime (\ref{dym}), we assume that the vector potentials $A_2,A_3$ vanishes and $A_0,A_1$ are functions of $(x_2,x_3)$. Therefore, $F_{01}=F_{23}=0$. The Maxwell's equations are
\begin{align}
    \nabla_{\lambda}F^{\alpha\lambda}&=J^{\alpha},\label{Max1}\\
    \nabla_{[\alpha}F_{\beta\lambda]}&=0,\label{Max2}
\end{align}
where $J^{\alpha}$ is the current vector. As the relationship $F_{\alpha\beta}=\partial_{\alpha}A_{\beta}-\partial_{\beta}A_{\alpha}$, the Eq. (\ref{Max2}) is satisfied automatically.

Let's consider the Maxwell's equation (\ref{Max1}) first. Since we are concerned with black hole solutions, we set the current vector to zero, $J^{\alpha}=(0,0,0,0)$. For $\alpha=0$, we obtain
\begin{align}
    (e^{\psi-\nu+\mu_3-\mu_2}&(F_{02}+qF_{12}))_{,2}\nonumber\\&+(e^{\psi-\nu+\mu_2-\mu_3}(F_{03}+qF_{13}))_{,3}=0.
\end{align}
As a simple solution, we define
\begin{align}
e^{\psi-\nu+\mu_3-\mu_2}(F_{02}+qF_{12})&=-B_{,3},\\
e^{\psi-\nu+\mu_2-\mu_3}(F_{03}+qF_{13})&=B_{,2},
\end{align}
where $B$ is a function of $(x_2,x_3)$. In the stationary axisymmetric space-time, there are two new unknowns $A_0,A_1$ for the electromagnetic field. We replace one of them with $B$, and the other one is still selected as $A=A_1$. For $\alpha=1$, the Eq. (\ref{Max1}) can be re-expressed as
\begin{equation}\label{max2}
     (e^{\nu-\psi+\mu_3-\mu_2}A_{,2})_{,2}+(e^{\nu-\psi+\mu_2-\mu_3}A_{,3})_{,3}=q_{,2}B_{,3}-q_{,3}B_{,2}.
\end{equation}
For Eq. (\ref{Max2}), there exist extra equation for $\alpha=0$
\begin{equation}\label{max1}
    (e^{\nu-\psi+\mu_3-\mu_2}B_{,2})_{,2}+(e^{\nu-\psi+\mu_2-\mu_3}B_{,3})_{,3}=q_{,3}A_{,2}-q_{,2}A_{,3}.
\end{equation}
When $\alpha=1$, Eq. (\ref{Max2}) is automatically satisfied. Finally, we obtain the constraint equations (\ref{max2}) and (\ref{max1}) for $A,B$.

Using $A,B$ to express the energy-momentum of electromagnetic fields (\ref{EMT}), we have
\begin{align}
T^{(m)}{}^0{}_0&=-T^{(m)}{}^1{}_1=-\frac{1}{8\pi}\bigg(e^{-2\psi-2\mu_2}(A^2_{,2}+B^2_{,2})\nonumber\\&+e^{-2\psi-2\mu_3}(A^2_{,3}+B^2_{,3})
\bigg),\label{Tm1}\\
T^{(m)}{}^2{}_2&=-T^{(m)}{}^3{}_3=\frac{1}{8\pi}\bigg(e^{-2\psi-2\mu_2}(A^2_{,2}+B^2_{,2})\nonumber\\&-e^{-2\psi-2\mu_3}(A^2_{,3}+B^2_{,3})\bigg),
\end{align}
\begin{align}
T^{(m)}{}^0{}_1&=\frac{1}{4\pi}e^{-(2\psi+\mu_2+\mu_3)}(B_{,2}A_{,3}-A_{,2}B_{,3}),\\
T^{(m)}{}^2{}_3&=\frac{1}{4\pi}e^{-(2\psi+\mu_2+\mu_3)}(A_{,2}A_{,3}+B_{,2}B_{,3}),\label{Tm2}
\end{align}
where the above expressions are the tetrad components $T^{(m)}{}^a{}_b=T^{(m)}{}^{\mu}{}_{\nu}e^a{}_{\mu}e^{\nu}{}_{b}$.

\section{The ans\"{a}tz of St\"{u}ckelberg fields $\phi^a$}
In the original dRGT model \cite{deRham2011}, the St\"{u}ckelberg field $\phi^a$ are chosen to be unitary gauge $\phi^a=x^{\alpha}\delta_{\alpha}^a$ and the reference metric is $f_{ab}=\eta_{ab}$. There exists a new basic invariant $I^{ab}=g^{\alpha\beta}\partial_{\alpha}\phi^a\partial_{\beta}\phi^b$ in massive gravity. When searching the black hole solution, $g^{\mu\nu}$ would possess coordinate singularity in the event horizon. In the unitary gauge, $I^{ab}$ would also possess singularity in the event horizon. The singularities in $I^{ab}$ of a black hole encounter a problem of fluctuations \cite{Berezhiani2011}. Thus, in order to obtain the healthy SSS black hole, we need to abandon the requirement of the unitary gauge. In the previous Ref. \cite{Li2016-1}, we use the nonunitary gauge to search for SSS solutions in dRGT theory. When serching the rotating black hole solution, one may also use the nonunitary gauge. However, instead of choosing the nonunitary gauge, we found a more concise approach.

Lemma: \emph{Arbitrary nonunitary gauge $\phi^a$ with the Minkowski reference metric $f_{ab}=\eta_{ab}$ would always led to the unitary gauge $\bar{\phi}^a=x^{\alpha}\delta_{\alpha}^a$ with some curved reference metric $\bar{f}_{ab}$.}

Proof: In the nonunitary gauge $\phi^a$ with the Minkowski reference metric $f_{ab}=\eta_{ab}$, there is $\gamma^{2}{}^{\alpha}{}_ {\beta}=g^{\alpha\sigma}\partial_{\sigma}\phi^a\partial_{\beta}\phi^b\eta_{ab}$. In the unitary gauge $\bar{\phi}^a=x^{\alpha}\delta_{\alpha}^a$ with a curved reference metric $\bar{f}_{ab}$, there is $\bar{\gamma}^{2}{}^{\alpha}{}_ {\beta}=g^{\alpha\sigma}\delta_{\sigma}^a\delta_{\beta}^b\bar{f}_{ab}$. When $\bar{f}_{\alpha\beta}=\partial_{\alpha}\phi^a\partial_{\beta}\phi^b\eta_{ab}$, the two matrix $\gamma^{2}$ derived by two different methods are equal $\gamma^{2}{}^{\alpha}{}_ {\beta}=\bar{\gamma}^{2}{}^{\alpha}{}_ {\beta}$. $\square$

Actually, we're more concerned with gravitational effects than we are with the St\"{u}ckelberg field $\phi^a$. The same matrix $\gamma^{2}$ would derive the same $T^{(\mathcal{K})}{}^{\alpha}{}_{\beta}$. Only by gravitational effects, we cannot tell whether the graviton mass term $m$ is provided by the St\"{u}ckelberg field $\phi^a$ or by the reference metric $\bar{f}_{ab}$. At the same time, the above lemma brings great convenience to the process of serching black hole solutions in dRGT model. Using the lemma, we choose the unitary gauge $\phi^a=x^{\alpha}\delta_{\alpha}^a$ of the St\"{u}ckelberg field but with a curved reference metric $f_{ab}$. Similar to the form (\ref{AxiForm}), the reference metric $f_{ab}$ can be chosen to
\begin{equation}\label{refM}
  f_{ab}=\left(
           \begin{array}{cccc}
             f_{00}(x^2 ,x^3) & f_{01}(x^2 ,x^3) & 0 & 0 \\
             f_{01}(x^2 ,x^3) & f_{11}(x^2 ,x^3) & 0 & 0 \\
             0 & 0 & f_{22}(x^2 ,x^3) & f_{23}(x^2 ,x^3) \\
             0 & 0 & f_{23}(x^2 ,x^3) & f_{33}(x^2 ,x^3) \\
           \end{array}
         \right)
  .
\end{equation}
Notice that the sub-metric $f_{ab}$ of the two-dimensional space $(x^2 ,x^3)$ is not an diagonal form. We are not sure if there is a connection between the dynamic metric $g_{\mu\nu}$ and the reference metric $f_{ab}$. Therefore, we did not rotate the two-dimensional space $(x^2 ,x^3)$ of the reference metric $f_{ab}$.

Taking the symmetry of the reference metric $f_{ab}=f_{ba}$, we actually introduce six unknowns $f_{00},f_{11},f_{22},f_{33},f_{01},f_{23}$. Using the reference metric (\ref{refM}) and the unitary gauge $\phi^a=x^{\alpha}\delta_{\alpha}^a$, one obtain
\begin{equation}\label{Ansatz}
  \gamma^{2}=\left(
                                 \begin{array}{cc}
                                  \Xi_1 & 0 \\
                                   0 & \Xi_2 \\
                                 \end{array}
                               \right) ,
\end{equation}
where $0$ is a $2\times 2$ zero matrix and
\begin{align}
  \Xi_1&=\left(
          \begin{array}{cc}
           -e^{-2\nu}(f_{00}+qf_{01})  & -e^{-2\nu}(f_{01}+qf_{11})  \\
           e^{-2\psi}f_{10}-e^{-2\nu}q(f_{00}+qf_{01})  & e^{-2\psi}f_{11}-e^{-2\nu}q(f_{01}+qf_{11}) \\
          \end{array}
        \right),\\
  \Xi_2&=\left(
          \begin{array}{cc}
           e^{-2\mu_2}f_{22} & e^{-2\mu_2}f_{23}  \\
           e^{-2\mu_3}f_{23}  &  e^{-2\mu_3}f_{33}\\
          \end{array}
        \right).
\end{align}
In fact, we require the matrix square root $\gamma$ to derive $T^{(\mathcal{K})}_{\alpha\beta}$. Notice that the matrix square root $\gamma$ can be expressed as
\begin{equation}
  \gamma=\left(
                                 \begin{array}{cc}
                                  \sqrt{\Xi_1} & 0 \\
                                   0 & \sqrt{\Xi_2} \\
                                 \end{array}
                               \right) .\label{GAMA}
\end{equation}
The square root of the $2\times 2$ submatrices $\sqrt{\Xi_1}$ and $\sqrt{\Xi_2}$ can be derived using the Cayley-Hamilton theorem.

The Cayley-Hamilton theorem: \emph{For a general $2\times 2$ matrix $M$, the Cayley-Hamilton theorem shows
\begin{equation}\label{CH}
  [M] M=M^2+(\text{det}~M)I_2,
\end{equation}
where $I_2$ is $2\times 2$ identity matrix.}

The Cayley-Hamilton theorem can be proved by direct calculation. We will not show the proof process in this paper. In addition to the Cayley-Hamilton theorem, we will also use the following properties. Let's say $M^2=\left(
           \begin{array}{cc}
             M^2_{11} & M^2_{12} \\
             M^2_{21} & M^2_{22} \\
           \end{array}
         \right)
$.
Taking the trace of Eq. (\ref{CH}), we have
\begin{equation}
  [M]^2=[M^2]+2\text{det}~M.
\end{equation}
 Using the property $(\text{det}~M)^n=\text{det}~M^n$, one obtains
\begin{align}
  [M]^2 & =M^2_{11}+ M^2_{22}+2\text{det}~M, \\
  (\text{det}~M)^2 & =M^2_{11} M^2_{22}-M^2_{12} M^2_{21}.
\end{align}
On the other hand, one can also use the trace $[M]$ and determinant $\text{det}~M$ to reexpress the diagonal component of $M^2$
\begin{align}
  M^2_{11} &= \frac{1}{2}\bigg([M]^2-2\text{det}~M\nonumber\\
  &-\sqrt{[M]^4-4[M]^2 \text{det}~M-4M^2_{12} M^2_{21}} \bigg),\label{M1} \\
  M^2_{22} &= \frac{1}{2}\bigg( [M]^2-2\text{det}~M\nonumber\\
  &+\sqrt{[M]^4-4[M]^2 \text{det}~M-4M^2_{12} M^2_{21}} \bigg).\label{M2}
\end{align}

The above properties imply that
\begin{align*}
  [\sqrt{\Xi_1}] & =\big(e^{-2\psi}f_{11}-e^{-2\nu}(f_{00}+2f_{01}q+f_{11}q^2)\\
  &-2 e^{-(\nu+\psi)} \sqrt{f_{01}^2-f_{00}f_{11}}\big)^{1/2},\\
  \text{det} ~\sqrt{\Xi_1} & =e^{-(\nu+\psi)} \sqrt{f_{01}^2-f_{00}f_{11}},\\
  [\sqrt{\Xi_2}] &=\big(e^{-2\mu_2}f_{22}+e^{-2\mu_3}f_{33}\\
  &-2e^{-\mu_2-\mu_3}\sqrt{f_{22}f_{33}-f_{23}^2}\big)^{1/2} ,\\
  \text{det}~ \sqrt{\Xi_2}&=e^{-\mu_2-\mu_3}\sqrt{f_{22}f_{33}-f_{23}^2} .
\end{align*}
Using the Cayley-Hamilton theorem (\ref{CH}), one can obtain the square root $\sqrt{\Xi_1}$ and $\sqrt{\Xi_2}$ as
\begin{align}
  \sqrt{\Xi_1} & =\frac{1}{[\sqrt{\Xi_1}]}\left(
                    \begin{array}{cc}
                      \frac{1}{2}( [\sqrt{\Xi_1}]^2-s_1)& -e^{-2\nu}(f_{01}+qf_{11}) \\
                      e^{-2\psi}f_{10}-e^{-2\nu}q(f_{00}+qf_{01}) & \frac{1}{2}( [\sqrt{\Xi_1}]^2+s_1)\\
                    \end{array}
                  \right),\label{x1}   \\
  \sqrt{\Xi_2} & =\frac{1}{ [\sqrt{\Xi_2}]}\left(
                    \begin{array}{cc}
                      \frac{1}{2}( [\sqrt{\Xi_2}]^2-s_2) & e^{-2\mu_2}f_{23} \\
                       e^{-2\mu_3}f_{23} & \frac{1}{2}( [\sqrt{\Xi_2}]^2+s_2) \\
                    \end{array}
                  \right),\label{x2}
\end{align}
where we have use Eqs. (\ref{M1}) and (\ref{M2}) to reexpress the diagonal components and
\begin{align}
  s_1&=\big( [\sqrt{\Xi_1}]^4-4 [\sqrt{\Xi_1}]^2 \text{det} ~\sqrt{\Xi_1}+ 4e^{-2\nu}(f_{01}\nonumber\\
  &+qf_{11}) (e^{-2\psi}f_{10}-e^{-2\nu}q(f_{00}+qf_{01}) )\big)^{1/2}, \\
  s_2&=\sqrt{ [\sqrt{\Xi_2}]^4-4 [\sqrt{\Xi_2}]^2 \text{det} ~\sqrt{\Xi_2}-4e^{-2(\mu_2+\mu_3)}f_{23}^2}.
\end{align}
Whin $\gamma$ (\ref{GAMA}), the expression of $T^{(\mathcal{K})}{}^{\alpha}{}_{\beta}$ is determined. For brevity, we define new constant
\begin{align}
  c_3 & =3\alpha_3+12\alpha_4, \\
  c_4 & =1+6\alpha_3+12\alpha_4.
\end{align}
Inserting Eqs. (\ref{GAMA}) (\ref{x1}) and (\ref{x2}) into (\ref{TK}), one obtains the tetrad components
\begin{align}
 T^{(\mathcal{K})}{}^{0}{}_{0} & =\frac{1}{2[\sqrt{\Xi_1}]}\bigg([\sqrt{\Xi_1}](-3 + 2 [\sqrt{\Xi_2}]- 2 c_3 ([\sqrt{\Xi_2}]-1)  \nonumber\\&+ c_4 (-3 - 2 \text{det} ~\sqrt{\Xi_2} + 4 [\sqrt{\Xi_2}]) )+([\sqrt{\Xi_1}]^2+s_1\nonumber\\&+2e^{-2\nu}(f_{01}+qf_{11})q)(1+ c_3(\text{det} ~\sqrt{\Xi_2}-1)\nonumber\\&+c_4 ( 2-[\sqrt{\Xi_2}])) \bigg),\label{TK1} \\
 T^{(\mathcal{K})}{}^{1}{}_{1}& =\frac{1}{2[\sqrt{\Xi_1}]}\bigg([\sqrt{\Xi_1}](-3 + 2 [\sqrt{\Xi_2}]- 2 c_3 ([\sqrt{\Xi_2}]-1)  \nonumber\\&+ c_4 (-3 - 2 \text{det} ~\sqrt{\Xi_2} + 4 [\sqrt{\Xi_2}]) )+([\sqrt{\Xi_1}]^2-s_1\nonumber\\&-2e^{-2\nu}(f_{01}+qf_{11})q)(1+ c_3(\text{det} ~\sqrt{\Xi_2}-1)\nonumber\\&+c_4 ( 2-[\sqrt{\Xi_2}]))\bigg), 
 \end{align}
 \begin{align}
 T^{(\mathcal{K})}{}^{2}{}_{2} & =\frac{1}{2[\sqrt{\Xi_2}]}\bigg([\sqrt{\Xi_2}](-3 + 2 [\sqrt{\Xi_1}]- 2 c_3 ([\sqrt{\Xi_1}]-1) \nonumber\\& + c_4 (-3 - 2 \text{det} ~\sqrt{\Xi_1} + 4 [\sqrt{\Xi_1}]) )+([\sqrt{\Xi_2}]^2\nonumber\\&+s_2)(1+ c_3(\text{det} ~\sqrt{\Xi_1}-1)+c_4 ( 2-[\sqrt{\Xi_1}])) \bigg),\\
 T^{(\mathcal{K})}{}^{3}{}_{3} & =\frac{1}{2[\sqrt{\Xi_2}]}\bigg([\sqrt{\Xi_2}](-3 + 2 [\sqrt{\Xi_1}]- 2 c_3 ([\sqrt{\Xi_1}]-1) \nonumber\\& + c_4 (-3 - 2 \text{det} ~\sqrt{\Xi_1} + 4 [\sqrt{\Xi_1}]) )+([\sqrt{\Xi_2}]^2\nonumber\\&-s_2)(1+ c_3(\text{det} ~\sqrt{\Xi_1}-1)+c_4 ( 2-[\sqrt{\Xi_1}])) \bigg),\\
 T^{(\mathcal{K})}{}^{0}{}_{1} & =\frac{(f_{01}+qf_{11})e^{-\nu-\psi}}{[\sqrt{\Xi_1}]}(1+ c_3(\text{det} ~\sqrt{\Xi_2}-1)\nonumber\\&+c_4 ( 2-[\sqrt{\Xi_2}])),\\
 T^{(\mathcal{K})}{}^{2}{}_{3} & =-\frac{f_{23}e^{-\mu_2-\mu_3}}{[\sqrt{\Xi_2}]}(1+ c_3(\text{det} ~\sqrt{\Xi_1}-1)\nonumber\\&+c_4 ( 2-[\sqrt{\Xi_1}])).\label{TK2}
\end{align}
\subsection{the cosmological constant term $\Lambda$}
The above expressions (\ref{TK1}) - (\ref{TK2}) of $T^{(\mathcal{K})}{}^{a}{}_{b}$ contain too many unknowns. To make the actual calculation more explicit, we need to simplify expressions further. In fact, the simplification is to set all or some of the unknown functions of $f_{ab}$. In the simplest case, $T^{(\mathcal{K})}{}^{a}{}_{b}$ is reduced to a cosmological constant term $\Lambda$. This case is denoted by $T^{(\mathcal{K})}_{\Lambda}{}^{a}{}_{b}$. Let's define
\begin{align}
   1+ c_3(\text{det} ~\sqrt{\Xi_2}-1)+c_4 ( 2-[\sqrt{\Xi_2}])&=0,\label{SolGama1}\\
   1+ c_3(\text{det} ~\sqrt{\Xi_1}-1)+c_4 ( 2-[\sqrt{\Xi_1}])&=0, \\
  -3 + 2 [\sqrt{\Xi_2}]- 2 c_3 ([\sqrt{\Xi_2}]-1) \nonumber\\ + c_4 (-3 - 2 \text{det} ~\sqrt{\Xi_2} + 4 [\sqrt{\Xi_2}])&=2\Lambda,\\
  -3 + 2 [\sqrt{\Xi_1}]- 2 c_3 ([\sqrt{\Xi_1}]-1) \nonumber\\ + c_4 (-3 - 2 \text{det} ~\sqrt{\Xi_1} + 4 [\sqrt{\Xi_1}])&=2\Lambda,\label{SolGama4}
\end{align}
where $\Lambda$ is a constant. These four equations show that
\begin{align}
\text{det} ~\sqrt{\Xi_1}&=\text{det} ~\sqrt{\Xi_2}=\nonumber\\&-\frac{3 c_3 - 2 c_3^2 - 2 c_4 + 5 c_3 c_4 - 4 c_4^2 + 2 c_3 \Lambda}{2 (-c_3 + c_3^2 - 2 c_3 c_4 + c_4^2)},\label{SolD}\\
[\sqrt{\Xi_1}] &=[\sqrt{\Xi_2}]=\nonumber\\&-\frac{-2 + 4 c_3 - 2 c_3^2 - 5 c_4 + 6 c_3 c_4 - 5 c_4^2 + 2 c_4\Lambda}{2 (-c_3 + c_3^2 - 2 c_3 c_4 + c_4^2)}.\label{SolT1}
\end{align}
Then, one can simplify
\begin{equation}\label{TLamda}
   T^{(\mathcal{K})}_{\Lambda}{}^{a}{}_{b}=\Lambda \delta^{a}_{b}.
\end{equation}

\subsection{the St\"{u}ckelberg charge $S$}
The result of simplification is not unique. By carefully comparing $T^{(\mathcal{K})}{}^{a}{}_{b}$ (\ref{TK1}) - (\ref{TK2}) and $T^{(m)}{}^{a}{}_{b}$ (\ref{Tm1}) - (\ref{Tm2}), we find that it is possible to simplify $T^{(\mathcal{K})}{}^{a}{}_{b}$ to a similiar configuration with  $T^{(m)}{}^{a}{}_{b}$ of the electromagnetic field. We denote this case by $T^{(\mathcal{K})}_S{}^{a}{}_{b}$. New charges can only be introduced by the St\"{u}ckelberg field. Thus, we call it the St\"{u}ckelberg charge $S$. Let's try to take $T^{(\mathcal{K})}{}^{a}{}_{b}$ as
\begin{align}
T^{(\mathcal{K})}_S{}^0{}_0&=-T^{(\mathcal{K})}_S{}^1{}_1=-e^{-2\psi-2\mu_2}(A'^2_{,2}+B'^2_{,2})\nonumber\\&-e^{-2\psi-2\mu_3}(A'^2_{,3}+B'^2_{,3}
),\label{TmS1}\\
T^{(\mathcal{K})}_S{}^2{}_2&=-T^{(\mathcal{K})}_S{}^3{}_3=e^{-2\psi-2\mu_2}(A'^2_{,2}+B'^2_{,2})\nonumber\\&-e^{-2\psi-2\mu_3}(A'^2_{,3}+B'^2_{,3}),\\
T^{(\mathcal{K})}_S{}^0{}_1&=2e^{-(2\psi+\mu_2+\mu_3)}(B'_{,2}A'_{,3}-A'_{,2}B'_{,3}),\\
T^{(\mathcal{K})}_S{}^2{}_3&=2e^{-(2\psi+\mu_2+\mu_3)}(A'_{,2}A'_{,3}+B'_{,2}B'_{,3}),\label{TmS2}
\end{align}
where $A',B'$ are the potential introduced by the St\"{u}ckelberg charge $S$ and have the similar properties with $A,B$. They satisfied a similar contraint equations for $A,B$, which are
\begin{align}
     (e^{\nu-\psi+\mu_3-\mu_2}A'_{,2})_{,2}+(e^{\nu-\psi+\mu_2-\mu_3}A'_{,3})_{,3}=q_{,2}B'_{,3}-q_{,3}B'_{,2},\label{StuckC1}\\
     (e^{\nu-\psi+\mu_3-\mu_2}B'_{,2})_{,2}+(e^{\nu-\psi+\mu_2-\mu_3}B'_{,3})_{,3}=q_{,3}A'_{,2}-q_{,2}A'_{,3}.\label{StuckC2}
\end{align}
The above two cases can be considered together. In order to facilitate uniform calculation, we actually consider $T^{(\mathcal{K})}{}^{a}{}_{b}=T^{(\mathcal{K})}_{\Lambda}{}^{a}{}_{b}+T^{(\mathcal{K})}_S{}^{a}{}_{b}$ in this paper.

\section{Gauge freedom and the Field equations}
Analytical calculation of the field equations (\ref{ModEin}) is not an easy task. The original mathematical process is detailed in Refs. \cite{Newman1965,Debney1969}. A more streamlined calculation process is provided by Carter \cite{Carter2009} or Chandrasekhar \cite{Chandrasekhar:1983}. In this paper, we employ the methodology primarily adopted from Chandrasekhar. As a key strategy of Chandrasekhar, we require a highly symmetric metirc form \cite{Li2024-2}
\begin{equation}\label{CForm}
  ds^2=\sqrt{\Delta_r\Delta_{\mu}}\left(-\chi dt^2+\frac{1}{\chi} (d\varphi-qdt)^2\right)+\rho^2\left(\frac{dr^2}{\Delta_r}+\frac{d\mu^2}{\Delta_{\mu}}\right),
\end{equation}
where $\chi=\chi(r,\mu),q=q(r,\mu)$. The function $\Delta_r$ depends only on $r$ and $\Delta_{\mu}$ depends only on $\mu$. The new coordinates $(r,\mu)$ are related to $(x^2,x^3)$ by the relationships
\begin{align}
  e^{2\mu_2(x^2,x^3)} &=\frac{\rho^2(r,\mu)}{\Delta_r} ,\\
  e^{2\mu_3(x^2,x^3)} & =\frac{\rho^2(r,\mu)}{\Delta_{\mu}}.
\end{align}
In the Boyer-Lindquist coordinate system, $r$ is the radial coordinate and $\mu=-\cos\theta$. The metric form (\ref{CForm}) can always be achieved by choosing the appropriate gauge freedom. Comparing (\ref{CForm}) with (\ref{dym}), we have
\begin{equation}\label{Metr1}
   e^{2\nu}  =\chi\sqrt{\Delta_r\Delta_{\mu}},~~e^{2\psi}  =\frac{\sqrt{\Delta_r\Delta_{\mu}}}{\chi},~~e^{2\mu_2}=\frac{\rho^2}{\Delta_r},~~ e^{2\mu_3}=\frac{\rho^2}{\Delta_{\mu}}.
\end{equation}
Notice that $(dx^3)^2=d\mu^2$, therefore $x^3=\pm \mu$. In the next calculation, we choose $x^3=-\mu$.

In this paper, we concern the solution of (\ref{ModEin}) with electromagnetic fields. Since the Bianchi identity $\nabla^{\alpha}G_{\alpha\beta}=0$ is satisfied, we consider the following 5 equations,
\begin{align}
 G^2{}_2+G^3{}_3+m^2(T^{(\mathcal{K})}{}^{2}{}_{2}+T^{(\mathcal{K})}{}^{3}{}_{3})  & =8\pi(T^{(m)}{}^{2}{}_{2}+T^{(m)}{}^{3}{}_{3}),\label{EinR1} \\
G^2{}_2-G^3{}_3+m^2(T^{(\mathcal{K})}{}^{2}{}_{2}-T^{(\mathcal{K})}{}^{3}{}_{3}) & =8\pi(T^{(m)}{}^{2}{}_{2}-T^{(m)}{}^{3}{}_{3}),\label{EinR11}\\
G^1{}_1+G^0{}_0+m^2(T^{(\mathcal{K})}{}^{1}{}_{1}+T^{(\mathcal{K})}{}^{0}{}_{0}) & =8\pi(T^{(m)}{}^{1}{}_{1}+T^{(m)}{}^{0}{}_{0}),\label{EinR21}\\
  G^1{}_1-G^0{}_0+m^2(T^{(\mathcal{K})}{}^{1}{}_{1}-T^{(\mathcal{K})}{}^{0}{}_{0}) & =8\pi(T^{(m)}{}^{1}{}_{1}-T^{(m)}{}^{0}{}_{0}),\label{EinR2}\\
  G^0{}_1+m^2T^{(\mathcal{K})}{}^{0}{}_{1}&=8\pi T^{(m)}{}^{0}{}_{1}.\label{EinR3}
\end{align}
Besides, there are two more contraint equations (\ref{max2}) and (\ref{max1}) for $A,B$. Pluging all expressions into (\ref{EinR1}) - (\ref{EinR3}), we obtain
\begin{align}
  \Delta_{r,rr}+ \Delta_{\mu,\mu\mu}+4m^2\Lambda \rho^2 =0&,\label{Eq1}  \\
  \Delta_{r,rr}- \Delta_{\mu,\mu\mu}-\frac{1}{4}\frac{\Delta_{r,r}^2}{\Delta_{r}}+\frac{1}{4}\frac{\Delta_{\mu,\mu}^2}{\Delta_{\mu}}-\Delta_{r,r}(\ln\rho^2)_{,r}&\nonumber\\+
  \Delta_{\mu,\mu}(\ln\rho^2)_{,\mu}
  +\frac{\Delta_{r}}{\chi^2}(\chi^2_{,r}-q^2_{,r})-\frac{\Delta_{\mu}}{\chi^2}(\chi^2_{,\mu}-q^2_{,\mu})&\nonumber\\
  +2\chi\bigg(\sqrt{\frac{\Delta_{\mu}}{\Delta_{r}}}(A_{,\mu}^2-m^2A'{}_{,\mu}^2+B_{,\mu}^2-m^2B'{}_{,\mu}^2)&\nonumber\\-\sqrt{\frac{\Delta_{r}}{\Delta_{\mu}}}(A_{,r}^2-m^2A'{}_{,r}^2+B_{,r}^2-m^2B'{}_{,r}^2)
  \bigg)=0&,\label{Eq11}\\
  \Delta_{r,rr}+ \Delta_{\mu,\mu\mu}-\frac{1}{4}\frac{\Delta_{r,r}^2}{\Delta_{r}}-\frac{1}{4}\frac{\Delta_{\mu,\mu}^2}{\Delta_{\mu}}
  +2\Delta_{r}(\ln \rho^2)_{,rr}&\nonumber\\+2\Delta_{\mu}(\ln \rho^2)_{,\mu\mu}
  +\Delta_{r,r}(\ln \rho^2)_{,r}+\Delta_{\mu,\mu}(\ln \rho^2)_{,\mu}&\nonumber\\+\frac{\Delta_{r}}{\chi^2}(\chi^2_{,r}-q^2_{,r})+\frac{\Delta_{\mu}}{\chi^2}(\chi^2_{,\mu}-q^2_{,\mu})+m^2\Lambda \rho^2=0&,\label{Eq21}\\
  \frac{1}{\chi^2}(\Delta_r q_{,r}^2+\Delta_{\mu} q_{,\mu}^2)  -\left(\frac{\Delta_r}{\chi}\chi_{,r}\right)_{,r}-\left(\frac{\Delta_{\mu}}{\chi}\chi_{,\mu}\right)_{,\mu}&\nonumber\\+
  2\chi\bigg(\sqrt{\frac{\Delta_{\mu}}{\Delta_{r}}}(A_{,\mu}^2-m^2A'{}_{,\mu}^2+B_{,\mu}^2-m^2B'{}_{,\mu}^2)&\nonumber\\+\sqrt{\frac{\Delta_{r}}{\Delta_{\mu}}}(A_{,r}^2-m^2A'{}_{,r}^2+B_{,r}^2-m^2B'{}_{,r}^2)
  \bigg)=0&,\label{Eq2}\\
  \left(\frac{\Delta_r}{\chi^2}q_{,r}\right)_{,r}+\left(\frac{\Delta_{\mu}}{\chi^2}q_{,\mu}\right)_{,\mu}-4\bigg(B_{,r}A_{,\mu}-B_{,\mu}A_{,r}&\nonumber\\+m^2B'_{,\mu}A'_{,r}- m^2B'_{,r}A'_{,\mu} \bigg)=0&.\label{Eq3}
\end{align}
The constraint equations for $A,B$ (\ref{max1}) and (\ref{max2}) can be expressed as
\begin{align}
    \left(\chi \sqrt{\frac{\Delta_{r}}{\Delta_{\mu}}}B_{,r}\right)_{,r}+
    \left(\chi \sqrt{\frac{\Delta_{\mu}}{\Delta_{r}}}B_{,\mu}\right)_{,\mu}
    &=q_{,r}A_{,\mu}-q_{,\mu}A_{,r},\label{Eq4}\\
    \left(\chi \sqrt{\frac{\Delta_{r}}{\Delta_{\mu}}}A_{,r}\right)_{,r}+
    \left(\chi \sqrt{\frac{\Delta_{\mu}}{\Delta_{r}}}A_{,\mu}\right)_{,\mu}
    &=q_{,\mu}B_{,r}-q_{,r}B_{,\mu}.\label{Eq5}
\end{align}
There are also similar constraint equations for $A',B'$, which are
\begin{align}
    \left(\chi \sqrt{\frac{\Delta_{r}}{\Delta_{\mu}}}B'_{,r}\right)_{,r}+
    \left(\chi \sqrt{\frac{\Delta_{\mu}}{\Delta_{r}}}B'_{,\mu}\right)_{,\mu}
    &=q_{,r}A'_{,\mu}-q_{,\mu}A'_{,r},\label{Eq6}\\
    \left(\chi \sqrt{\frac{\Delta_{r}}{\Delta_{\mu}}}A'_{,r}\right)_{,r}+
    \left(\chi \sqrt{\frac{\Delta_{\mu}}{\Delta_{r}}}A'_{,\mu}\right)_{,\mu}
    &=q_{,\mu}B'_{,r}-q_{,r}B'_{,\mu}.\label{Eq7}
\end{align}
The equations above indicate that $A,B$ and $A',B'$ always appear in pairs. We also found that the corner marks $r,\mu$ appear in pairs, too. This is a key point to solve them.

\section{the Ernst equations}
Similar to Ref. \cite{Li2024-2}, we solve Eqs. (\ref{Eq1})  and (\ref{Eq2}) - (\ref{Eq7}) simultaneously. We rewrite the pairs $A,B$ into a single complex potential
\begin{equation}\label{ComplexPotential1}
 H=A+i B.
\end{equation}
The constraint Eqs. (\ref{Eq4}) and (\ref{Eq5}) can be combined to a single equation
\begin{equation}
      \left(\chi \sqrt{\frac{\Delta_{r}}{\Delta_{\mu}}}H_{,r}\right)_{,r}+
    \left(\chi \sqrt{\frac{\Delta_{\mu}}{\Delta_{r}}}H_{,\mu}\right)_{,\mu}
    =i(q_{,r}H_{,\mu}-q_{,\mu}H_{,r}).\label{ComplexEq1}
\end{equation}
For the the pairs $A',B'$, we also define $V=A'+i B'$ and derive the similar equation
\begin{equation}
      \left(\chi \sqrt{\frac{\Delta_{r}}{\Delta_{\mu}}} V_{,r}\right)_{,r}+
    \left(\chi \sqrt{\frac{\Delta_{\mu}}{\Delta_{r}}}V_{,\mu}\right)_{,\mu}
    =i(q_{,r}V_{,\mu}-q_{,\mu}V_{,r}).\label{ComplexEq2}
\end{equation}
Notice that there are
\begin{align}
  H H^{*}_{,r} & =AA_{,r}+BB_{,r}+i(BA_{,r}-AB_{,r}), \\
  |H_{,r}|^2&=A_{,r}^2+B_{,r}^2,
\end{align}
where $*$ denotes the conjugate of complex numbers and $||$ denotes the modular length of a complex number. Then, we can reexpressed Eqs. (\ref{Eq2}) and (\ref{Eq3}) as
\begin{align}
&(\Delta_{r}(\ln \chi)_{,r})_{,r}+ (\Delta_{r}(\ln \chi)_{,\mu})_{,\mu}-\frac{1}{\chi^2}(\Delta_{r}q_{,r}^2+\Delta_{\mu}q_{,\mu}^2) =  \nonumber\\&
\frac{2\chi}{\sqrt{\Delta_r\Delta_{\mu}}}(\Delta_r(|H_{,r}|^2-m^2|V_{,r}|^2)+\Delta_{\mu}(|H_{,\mu}|^2-m^2|V_{,\mu}|^2))&     \label{CEq1}\\
  &\left(\frac{\Delta_r}{\chi^2}q_{,r}-2\text{Im}(H H^{*}_{,\mu}-m^2V V^{*}_{,\mu})\right)_{,r}\nonumber\\&
  +\left(\frac{\Delta_{\mu}}{\chi^2}q_{,\mu}+2\text{Im}(H H^{*}_{,r}-m^2V V^{*}_{,r}) \right)_{,\mu}=0,\label{CEq2}
\end{align}
where $\text{Im}$ denotes the imaginary part. Eq. (\ref{CEq2}) permits a simple solution
\begin{align}
  -\Phi_{,r} & =\frac{\Delta_{\mu}}{\chi^2}q_{,\mu}+2\text{Im}(H H^{*}_{,r}-m^2V V^{*}_{,r}), \label{qr}\\
  \Phi_{,\mu} & =\frac{\Delta_r}{\chi^2}q_{,r}-2\text{Im}(H H^{*}_{,\mu}-m^2V V^{*}_{,\mu}),\label{qmu}
\end{align}
which is shown by $\Phi_{,\mu,r}=\Phi_{,r,\mu}$. At the same time, the equation $q_{,r,\mu}=q_{,\mu,r}$ requires that
\begin{align}
  & \left(\frac{\chi^2}{\Delta_{\mu}}(\Phi_{,r}+2\text{Im}(H H^{*}_{,r}-m^2V V^{*}_{,r})) \right)_{,r}\nonumber\\
  &+\left(\frac{\chi^2}{\Delta_{r}} (\Phi_{,\mu}+2\text{Im}(H H^{*}_{,\mu}-m^2V V^{*}_{,\mu}))\right)_{,\mu}=0.  \label{equation1}
\end{align}

Let's define
\begin{equation}
  \Psi=\frac{\sqrt{\Delta_{r}\Delta_{\mu}}}{\chi}.
\end{equation}
We can derive the following equations
\begin{align}
   & \Psi\{[\Delta_r(\Psi+|H|^2-m^2|V|^2 )_{,r}]_{,r} +[\Delta_{\mu}(\Psi+|H|^2\nonumber \\&-m^2|V|^2 )_{,\mu}]_{,\mu} \}=\frac{1}{2}\Psi^2(\Delta_{r,rr}+\Delta_{\mu,\mu\mu})  +\Delta_{r}\{\Psi_{,r}(\Psi+|H|^2\nonumber \\&-m^2|V|^2 )_{,r}-\Phi_{,r}^2-2\Phi_{,r}\text{Im}(H H^{*}_{,r}-m^2V V^{*}_{,r}) \}\nonumber \\
   &+\Delta_{\mu}\{\Psi_{,\mu}(\Psi+|H|^2-m^2|V|^2 )_{,\mu}\nonumber \\
   &-\Phi_{,\mu}^2-2\Phi_{,\mu}\text{Im}(H H^{*}_{,\mu}-m^2V V^{*}_{,\mu}) \},\label{E1}\\
   &\Psi[(\Delta_{r}\Phi_{,r} )_{,r}+ (\Delta_{\mu}\Phi_{,\mu})_{,\mu}] =\Delta_r\{\Phi_{,r}(2\Psi+|H|^2-m^2|V|^2)_{,r} \nonumber \\
  &+2(\Psi+|H|^2-m^2|V|^2)_{,r}\text{Im}(HH^{*}_{,r}-VV^{*}_{,r}) \}\nonumber \\
  & +\Delta_{\mu}\{\Phi_{,\mu}(2\Psi+|H|^2-m^2|V|^2)_{,\mu} \nonumber \\
  &+2(\Psi+|H|^2-m^2|V|^2)_{,\mu}\text{Im}(HH^{*}_{,\mu}-VV^{*}_{,\mu}) \},\label{E2}
\end{align}
where we have using the Eqs. (\ref{I}) (\ref{II}) (\ref{III}) (\ref{IV}) (\ref{V}) (\ref{VI}) in the appendix. Please refer to the appendix for the specific derivation process for these equations. Besides, using (\ref{qr1}) and (\ref{qmu1}), we can rewrite the Eq. (\ref{ComplexEq1}) as
\begin{align}
 & \Psi[(\Delta_{r}H_{,r})_{,r}+(\Delta_{\mu}H_{,\mu})_{,\mu}]=\Delta_r H_{,r}[\Psi_{,r}+i(\Phi_{,r}\nonumber\\
  &+2\text{Im}(HH^{*}_{,r}-m^2VV^{*}_{,r}) )]+\Delta_{\mu} H_{,\mu}[\Psi_{,\mu}+i(\Phi_{,\mu}\nonumber\\
  &+2\text{Im}(HH^{*}_{,\mu}-m^2VV^{*}_{,\mu}) )].\label{E3}
\end{align}
For $V$, there is
\begin{align}
 & \Psi[(\Delta_{r}V_{,r})_{,r}+(\Delta_{\mu}V_{,\mu})_{,\mu}]=\Delta_r V_{,r}[\Psi_{,r}+i(\Phi_{,r}\nonumber\\
  &+2\text{Im}(HH^{*}_{,r}-m^2VV^{*}_{,r}) )]+\Delta_{\mu} V_{,\mu}[\Psi_{,\mu}\nonumber\\
  &+i(\Phi_{,\mu}+2\text{Im}(HH^{*}_{,\mu}-m^2VV^{*}_{,\mu}) )].\label{E4}
\end{align}

We then define the complex function
\begin{equation}
  Z=\Psi+|H|^2-m^2|V|^2+i\Phi.
\end{equation}
We observe that in view of the identity
\begin{equation}
  2i \text{Im} HH^{*}_{,r}=|H|^2_{,r}-2H^{*}H_{,r},
\end{equation}
Eqs. (\ref{E1}) and (\ref{E2}) can be combined into a single expression
\begin{align}
 & \Psi[(\Delta_rZ_{,r})_{,r}+(\Delta_{\mu}Z_{,\mu})_{,\mu} ] =  \frac{1}{2}\Psi^2(\Delta_{r,rr}+\Delta_{\mu,\mu\mu})\nonumber \\
   &+\Delta_{r}Z_{,r}[Z_{,r}-2(H^{*}H_{,r}-m^2V^{*}V_{,r})]  \nonumber \\
   &+\Delta_{\mu}Z_{,\mu}[Z_{,\mu}-2(H^{*}H_{,\mu}-m^2V^{*}V_{,\mu})],\label{Ernst1}
\end{align}
where
\begin{equation}
  \Psi=\text{Re} Z-|H|^2+m^2|V|^2.
\end{equation}
The symbol $\text{Re}$ means the real part of a complex number. The other two equations (\ref{E3}) and (\ref{E4}) can be expressed as
\begin{align}
  &\Psi[(\Delta_rH_{,r})_{,r}+(\Delta_{\mu}H_{,\mu})_{,\mu} ] = \Delta_{r}H_{,r}[Z_{,r}-2(H^{*}H_{,r}-m^2V^{*}V_{,r})]  \nonumber \\
   & +\Delta_{\mu}H_{,\mu}[Z_{,\mu}-2(H^{*}H_{,\mu}-m^2V^{*}V_{,\mu})],\label{Ernst2}\\
   & \Psi[(\Delta_rV_{,r})_{,r}+(\Delta_{\mu}V_{,\mu})_{,\mu} ] = \Delta_{r}V_{,r}[Z_{,r}-2(H^{*}H_{,r}-m^2V^{*}V_{,r})]  \nonumber \\
   & +\Delta_{\mu}V_{,\mu}[Z_{,\mu}-2(H^{*}H_{,\mu}-m^2V^{*}V_{,\mu})].\label{Ernst3}
\end{align}
Eqs. (\ref{Eq1}) and (\ref{Ernst1}) - (\ref{Ernst3}) are called the Ernst equations.

\section{conjugate transformation}
The simplest rout to obtain the black hole solutions is transform the variables in a conjugate metric. Considering the metric transformation
\begin{equation}
  t\rightarrow i\varphi,\quad \varphi \rightarrow it,
\end{equation}
one can transform the metric into a conjugate form
\begin{align}
  \chi dt^2-\frac{1}{\chi}(d\varphi-q dt)^2\rightarrow & \chi dt^2+\frac{2q}{\chi}dtd\varphi-\frac{\chi^2-q^2}{\chi}d\varphi^2 \nonumber \\
   =&-\tilde{\chi} dt^2+\frac{1}{\tilde{\chi}}(d\varphi-\tilde{q} dt)^2,\nonumber
\end{align}
where
\begin{equation}
  \tilde{\chi}=-\frac{\chi}{\chi^2-q^2},~~~\tilde{q}=\frac{q}{\chi^2-q^2}.
\end{equation}
Eq. (\ref{ComplexEq1}) can be rewriten as the form
\begin{equation}
      \left(\chi \sqrt{\frac{\Delta_{r}}{\Delta_{\mu}}}H_{,r}-iq H_{,\mu}\right)_{,r}+
    \left(\chi \sqrt{\frac{\Delta_{\mu}}{\Delta_{r}}}H_{,\mu}+ iq H_{,r}\right)_{,\mu}
    =0.
\end{equation}
We can define a potential $\tilde{H}$ by the equations
\begin{align}
  -\tilde{H}_{,\mu} &=\chi \sqrt{\frac{\Delta_{r}}{\Delta_{\mu}}}H_{,r}-iq H_{,\mu},  \\
  \tilde{H}_{,r} & =\chi \sqrt{\frac{\Delta_{\mu}}{\Delta_{r}}}H_{,\mu}+ iq H_{,r}.
\end{align}
The above equations reduce to $\tilde{H}_{,\mu,r}=\tilde{H}_{,r,\mu}$. Solving these equations for $H_{,r}$ and $H_{,\mu}$, one have
\begin{align}
  H_{,r} &=\tilde{\chi} \sqrt{\frac{\Delta_{\mu}}{\Delta_{r}}}\tilde{H}_{,\mu}+i\tilde{q} \tilde{H}_{,r}, \label{exp1} \\
 -H_{,\mu} & =\tilde{\chi} \sqrt{\frac{\Delta_{r}}{\Delta_{\mu}}}\tilde{H}_{,r}- i\tilde{q} \tilde{H}_{,\mu}.\label{exp2}
\end{align}
Then, the requirement $H_{,\mu,r}=H_{,r,\mu}$ shows that
\begin{equation}
\left(\tilde{\chi} \sqrt{\frac{\Delta_{r}}{\Delta_{\mu}}}\tilde{H}_{,r}-i\tilde{q} \tilde{H}_{,\mu}\right)_{,r}+
    \left(\tilde{\chi} \sqrt{\frac{\Delta_{\mu}}{\Delta_{r}}}\tilde{H}_{,\mu}+ i\tilde{q} \tilde{H}_{,r}\right)_{,\mu}
    =0.
\end{equation}
We also express $\tilde{H}=\tilde{A}+i \tilde{B}$. Eqs. (\ref{exp1}) and (\ref{exp2}) indicate that
\begin{align}
  \tilde{A}_{,r} =\chi\sqrt{\frac{\Delta_{\mu}}{\Delta_{r}}}A_{,\mu}-q B_{,r},    \quad & \tilde{A}_{,\mu} =-\chi\sqrt{\frac{\Delta_{r}}{\Delta_{\mu}}}A_{,r}-q B_{,\mu}, \\
   \tilde{B}_{,r} =\chi\sqrt{\frac{\Delta_{\mu}}{\Delta_{r}}}B_{,\mu}+q A_{,r},    \quad & \tilde{B}_{,\mu} =-\chi\sqrt{\frac{\Delta_{r}}{\Delta_{\mu}}}B_{,r}+q A_{,\mu}.
\end{align}
Similarly, for $V$, we can also denfine $\tilde{V}=\tilde{A}'+i \tilde{B}'$, where $\tilde{A}',\tilde{B}'$ satify the same form equations for $\tilde{A},\tilde{B}$.

The directly calculations show that
\begin{align}
&\frac{1}{\tilde{\chi}^2}(\Delta_r \tilde{q}_{,r}^2+\Delta_{\mu} \tilde{q}_{,\mu}^2)  -\left(\frac{\Delta_r}{\tilde{\chi}}\tilde{\chi}_{,r}\right)_{,r}-\left(\frac{\Delta_{\mu}}{\tilde{\chi}}\tilde{\chi}_{,\mu}\right)_{,\mu}\nonumber\\&+
  2\tilde{\chi}\bigg(\sqrt{\frac{\Delta_{\mu}}{\Delta_{r}}}(\tilde{A}_{,\mu}^2-m^2\tilde{A}'{}_{,\mu}^2+\tilde{B}_{,\mu}^2-m^2\tilde{B}'{}_{,\mu}^2)\nonumber\\&+\sqrt{\frac{\Delta_{r}}{\Delta_{\mu}}}(\tilde{A}_{,r}^2-m^2\tilde{A}'{}_{,r}^2+\tilde{B}_{,r}^2-m^2\tilde{B}'{}_{,r}^2)
  \bigg)=0,\nonumber\\
 & \left(\frac{\Delta_r}{\tilde{\chi}^2}\tilde{q}_{,r}\right)_{,r}+\left(\frac{\Delta_{\mu}}{\tilde{\chi}^2}\tilde{q}_{,\mu}\right)_{,\mu}-4\bigg(\tilde{B}_{,r}\tilde{A}_{,\mu}\nonumber\\
 &-\tilde{B}_{,\mu}\tilde{A}_{,r}+m^2\tilde{B}'_{,\mu}\tilde{A}'_{,r}- m^2\tilde{B}'_{,r}\tilde{A}'_{,\mu} \bigg)=0,\nonumber
\end{align}
which are the exactly the form of Eqs. (\ref{Eq2}) and (\ref{Eq3}). From the fact that $\tilde{\chi},\tilde{q},\tilde{H},\tilde{V}$ statify the same equations as the "untilded" function, it follows that with further definitions,
\begin{align}
  \tilde{\Psi} &=\frac{\sqrt{\Delta_{r}\Delta_{\mu}}}{\tilde{\chi}},\label{psit}  \\
    -\tilde{\Phi}_{,r} & =\frac{\Delta_{\mu}}{\tilde{\chi}^2}\tilde{q}_{,\mu}-2\text{Im}(\tilde{H} \tilde{H}^{*}_{,r}-m^2\tilde{V} \tilde{V}^{*}_{,r}), \label{qrtild}\\
  \tilde{\Phi}_{,\mu} & =\frac{\Delta_r}{\tilde{\chi}^2}\tilde{q}_{,r}+2\text{Im}(\tilde{H} \tilde{H}^{*}_{,\mu}-m^2\tilde{V} \tilde{V}^{*}_{,\mu}),\label{qmutild}
\end{align}
and the functions
\begin{equation}
  \tilde{Z}=\tilde{\Psi}+|\tilde{H}|^2-m^2|\tilde{V}|^2+i\tilde{\Phi},
\end{equation}
will satisfy Ernst's equations of exactly the same forms as Eqs. (\ref{Ernst1}) - (\ref{Ernst3}).

\section{the homogeneous solution}
Similar to the Ref. \cite{Li2024-2}, we first solve the Eqs. (\ref{Eq1}) and (\ref{Ernst1}) - (\ref{Ernst3}) in the homogeneous case, which is $\Lambda=0$. We denote the homogeneous solution with a corner "K". When difine
\begin{equation}
  H^K=Q(Z^K+1),\quad V^K=S(Z^K+1),
\end{equation}
we found three Eqs. (\ref{Ernst1}) - (\ref{Ernst3}) reduce to the same equation
\begin{align}
 & \{ [\frac{1}{2}-(|Q|^2-m^2|S|^2)](Z^K+(Z^K)^{*})\nonumber\\
 &-(|Q|^2-m^2|S|^2)(|Z^K|^2+1)\}[(\Delta_r^K(Z^K)_{,r})_{,r}+(\Delta_{\mu}^K(Z^K)_{,\mu})_{,\mu}]\nonumber\\
 &= \{1-2(|Q|^2-m^2|S|^2)((Z^K)^{*}+1) \}[\Delta_r^K(Z^K)_{,r}^2+ \Delta_{\mu}^K(Z^K)_{,\mu}^2],\label{Zeq1}
\end{align}
where $Q$ and $S$ are complex number and we have use the identity $|Z+1|^2=|Z|^2+1+Z+Z^{*}$. In order to seek the solution, we make the further transformation
\begin{equation}
  Z^K=\frac{1+E^K}{1-E^K}.
\end{equation}
The direct calculation shows that
\begin{equation}
  \text{Re} Z^K=\frac{1-|E^K|^2}{|1-E^K|^2},\quad \text{Im} Z^K=-i\frac{E^K-(E^K)^{*}}{|1-E^K|^2}.
\end{equation}
Therefore, there is
\begin{align}
  \Psi^K&=\frac{1}{|1-E^K|^2}\{1-4(|Q|^2-m^2|S|^2)-|E^K|^2 \} ,\label{PSI1} \\
  \Phi^K&=-i\frac{E^K-(E^K)^{*}}{|1-E^K|^2}.
\end{align}
Eq. (\ref{Zeq1}) can be reexpressed as
\begin{align}
  &  \{1-4(|Q|^2-m^2|S|^2)-|E^K|^2 \}[(\Delta_r^KE^K_{,r})_{,r}+(\Delta_{\mu}^KE^K_{,\mu})_{,\mu}]\nonumber\\
  &=-2(E^K)^{*}[\Delta_r^K(E^K_{,r})^2+ \Delta_{\mu}^K(E^K_{,\mu})^2].\label{Zeq2}  
\end{align}

We gave the elementary solution of Eqs. (\ref{Eq1}) and (\ref{Zeq2}) in the case of $\Lambda=0$. At the first sight, one may think this is impossible. There are only two equations, but there are three unknowns $\Delta_{r},\Delta_{\mu},E$. We actully find the corner $r$ and $\mu$ always appear in pairs, which indicates that the dependence of $\Delta_r$ on $r$ is similar to the dependence of $\Delta_{\mu}$ on $\mu$. It is not difficult to show that $\Delta_r^K$ is a quadratic function of $r$. Similarly, $\Delta_{\mu}^K$ is a quadratic function of $\mu$. Then, we consider the conjugate function $\tilde{E}^K$ for the solution of homogeneous Ernst equations; and we can verify that the Ernst equations permit the elementary solution
\begin{align}
   \Delta_{r}^K&=r^2-1,  \label{SolK1}\\
   \Delta_{\mu}^K&=1-\mu^2,\label{SolK2}\\
   \tilde{E}^K&=-C_r r-i C_{\mu} \mu,
\end{align}
where $C_r$ and $C_{\mu}$ are real constants related in the manner
\begin{equation}
  C_r^2+C_{\mu}^2=1-4(|Q|^2-m^2|S|^2).\label{relation}
\end{equation}

They are not the unique solutions of the Ernst equations. For any quadratic functions
\begin{align}
 \Delta^K_{r}&=r^2+2C_1 r+D_1,  \\
  \Delta^K_{\mu}&=-(\mu^2+2C_2\mu+D_2),
\end{align}
where $C_1,D_1,C_2,D_2$ are real constants, we can obtain similar solutions. Defining
\begin{align}
  u & =\frac{r+C_1}{\sqrt{C_1^2-D_1}}, \\
  v & =\frac{\mu+C_2}{\sqrt{C_2^2-D_2}},
\end{align}
we have
\begin{align}
  \Delta_{r}^K &=(C_1^2-D_1)(u^2-1)\equiv(C_1^2-D_1)\Delta_u^K,  \\
  \Delta_{\mu}^K & =(C_2^2-D_2)(1-v^2)\equiv(C_2^2-D_2)\Delta_v^K.
\end{align}
Direct mathematical calculations show that $\Delta^K_u(u),\Delta^K_v(v),\tilde{E}^K(u,v)$ also satisfy the Ernst's equations, i.e.
\begin{align}
  \Delta^K_{u,uu}+\Delta^K_{v,vv} & =0,\nonumber \\
  (1-\tilde{E}^K(\tilde{E}^K)^{*})[(\Delta^K_u \tilde{E}^K_{,u})_{,u}+(\Delta^K_{v}\tilde{E}^K_{,v})_{,v}]&=\nonumber \\-2(\tilde{E}^K)^{*}[\Delta^K_u(\tilde{E}^K_{,u})^2+\Delta^K_{v}(\tilde{E}^K_{,v})^2]&.\nonumber
\end{align}
In virtue of these processes, we can obtain many new solutions of the Ernst equations.

The following solution is always ture
\begin{equation}
 \tilde{E}^K=-C_u u-i C_v v,
\end{equation}
where $C_u^2+C_v^2=1-4(|Q|^2-m^2|S|^2)$. Since $\tilde{Z}^K=\frac{1+\tilde{E}^K}{1-\tilde{E}^K}$, we have
\begin{align}
  \text{Re} \tilde{Z}^K & =-\frac{C_u^2(u^2-1)-C_v^2(1-v^2)-4(|Q|^2-m^2|S|^2)}{(1+C_u u)^2+C_v^2 v^2}, \\
  \text{Im} \tilde{Z}^K & =-\frac{2 C_v v}{(1+C_u u)^2+C_v^2 v^2}.
\end{align}
By Eq. (\ref{PSI1}), one have
\begin{equation}
  \tilde{\Psi}^K=-\frac{C_u^2(u^2-1)-C_v^2(1-v^2)}{(1+C_u u)^2+C_v^2 v^2}.
\end{equation}
Because of the similarity of $\tilde{H}$ and $\tilde{V}$, we choose
\begin{align}
  (\tilde{H}^K)^2 & =\frac{4|Q|^2}{(1+C_u u)^2+C_v^2 v^2}, \\
  (\tilde{V}^K)^2 & =\frac{4m^2|S|^2}{(1+C_u u)^2+C_v^2 v^2}.
\end{align}

Let's further choose
\begin{equation}
  u=\frac{r-M}{\sqrt{M^2-M_0^2}},\quad v=\mu.
\end{equation}
Thus, the expression
\begin{align}
   (1+C_u u)^2+C_v^2 v^2&=\frac{C_u^2}{M^2-M_0^2}\bigg[ \left(\frac{\sqrt{M-M_0}}{C_u}+r-M \right)^2\nonumber\\
   &+\frac{C_v^2}{C_u^2}(M^2-M_0^2)\mu^2 \bigg]   
\end{align}
is much simplied by the choice
\begin{equation}
  C_{u}=\frac{\sqrt{M^2-M_0^2}}{M},\quad C_v=\frac{a}{M},
\end{equation}
where $a$ is a constant. With this choice, one have $(1+C_u u)^2+C_v^2 v^2=M^2(r^2+a^2\mu^2)$. However, to be consistent with the relation (\ref{relation}), we must require that
\begin{equation}
  (M^2-M_0^2)+a^2=M^2[1-4(|Q|^2-m^2|S|^2)].
\end{equation}
Then,
\begin{equation}
  M_0=a^2+4M^2(|Q|^2-m^2|S|^2)\equiv a^2+Q_{*}^2-m^2S_{*}^2,
\end{equation}
where $Q_{*}^2=4M^2|Q|^2, S_{*}^2=4M^2|S|^2$.

To sum up, the homogeneous solutions are
\begin{align}
   \Delta_{r}^K&=r^2-2Mr+a^2+Q_{*}^2-m^2S_{*}^2,  \\
   \Delta_{\mu}^K&=1-\mu^2,\\
   \tilde{E}^K&=-C_r \frac{r-M}{\sqrt{M^2-a^2-Q_{*}^2+m^2S_{*}^2}}-i C_{\mu} \mu,
\end{align}
where $C_r=\frac{\sqrt{M^2-a^2-Q_{*}^2+m^2S_{*}^2}}{M}$ and $C_{\mu}=\frac{a}{M}$. After some tedious calculation, we express
\begin{align}
  \tilde{\Psi}^K & =-\frac{\Delta^K_r-a^2\Delta^K_{\mu}}{r^2+a^2\mu^2},\label{SolKK1} \\
  \tilde{\Phi}^K & =-\frac{2aM\mu}{r^2+a^2\mu^2},\\
  \tilde{H}^K&=Q_{*}\frac{r-ia\mu}{r^2+a^2\mu^2},\\
  \tilde{V}^K&=S_{*}\frac{r-ia\mu}{r^2+a^2\mu^2}.
\end{align}
Transformation into the orignal metric, the unknown functions are
\begin{align}
\rho^2&=r^2+a^2\mu^2,\label{rhoK}\\
  \chi & =\frac{(r^2+a^2\mu^2)\sqrt{\Delta^K_r\Delta^K_{\mu}}}{(r^2+a^2)^2\Delta^K_{\mu}-a^2(1-\mu^2)^2\Delta^K_r},\label{chiK} \\
  q &=a\frac{(r^2+a^2)\Delta^K_{\mu}-(1-\mu^2)\Delta^K_r}{(r^2+a^2)^2\Delta^K_{\mu}-a^2(1-\mu^2)^2\Delta^K_r},\label{qK}
\end{align}
and $A=-Q_{*}\frac{ra(1-\mu^2)}{\rho^2},B=-Q_{*}\frac{\mu(r^2+a^2)}{\rho^2}$ and $A'=-S_{*}\frac{ra(1-\mu^2)}{\rho^2},B'=-S_{*}\frac{\mu(r^2+a^2)}{\rho^2}$. If one choosing $\mu=-\cos\theta$, the homogeneous solutions can be expressed in the Boyer-Lindquist coordinate system
\begin{align}
  ds^2=&-\frac{\rho^2\Delta^K_r\Delta^K_{\theta}}{\Sigma^2}dt^2
  +\frac{\rho^2}{\Delta^K_r}dr^2+\frac{\rho^2\sin^2\theta}{\Delta^K_{\theta}}d\theta^2
  \nonumber\\
  &+\frac{\Sigma^2}{\rho^2}\bigg(d\varphi-\frac{(r^2+a^2)\Delta^K_{\theta}-\sin^2\theta \Delta^K_{r}}{\Sigma^2}adt\bigg)^2,\label{BLform1}
\end{align}
where
\begin{align}
  \rho^2 & =r^2+a^2\cos^2\theta, \\
  \Sigma^2 & =(r^2+a^2)^2\Delta^K_{\theta}-a^2\sin^4\theta\Delta^K_{r},\\
  \Delta^K_{r}&=r^2-2Mr+a^2+Q_{*}^2-m^2S_{*}^2,\\
  \Delta^K_{\theta}&=\sin^2\theta.
\end{align}

\section{the non-homogeneous solution}
In the non-homogeneous case, we expect that the soluton is linear depend on $\Lambda$. In the non-rotating case $a=0$, this can be easyly proof. Since the parameter $a$ and $\Lambda$ are independent, this algebraic is always ture. Then, we decompose the non-homogeneous solutions into two parts
\begin{align}
  \Delta_{r} & = \Delta_{r}^{K}+m^2\Lambda \Delta_{r}^{\Lambda},\label{Sol1}\\
  \Delta_{\mu} & = \Delta_{\mu}^{K}+m^2\Lambda \Delta_{\mu}^{\Lambda},\label{Sol2}\\
\tilde{Z}&=\tilde{Z}^K+m^2\Lambda \tilde{Z}^{\Lambda},
\end{align}
where $\Delta_{r}^{K},\Delta_{\mu}^{K},\tilde{Z}^K$ are homogeneous solutions and $\Delta_{r}^{\Lambda},\Delta_{\mu}^{\Lambda},\tilde{Z}^{\Lambda}$ are the solutions introduced by $\Lambda$. Notice that
\begin{equation}
  \Delta_{r}^{\Lambda}=\frac{\partial \Delta_{r}}{\partial (m^2\Lambda)},~\Delta_{\mu}^{\Lambda}=\frac{\partial \Delta_{\mu}}{\partial (m^2\Lambda)},~\tilde{Z}^{\Lambda}=\frac{\partial \tilde{Z}}{\partial (m^2\Lambda)}.\nonumber
\end{equation}
If one take the partial differential of $m^2\Lambda$, Eqs. (\ref{Eq1}) and (\ref{Ernst1}) can be rewriten as
\begin{align}
\Delta_{r,rr}^{\Lambda}+\Delta_{\mu,\mu\mu}^{\Lambda}&=-4(r^2+a^2\mu^2),\\
  \Psi^{\Lambda}[(\Delta_r^{\Lambda}\tilde{Z}^{\Lambda}_{,r})_{,r}+(\Delta_{\mu}^{\Lambda}\tilde{Z}^{\Lambda}_{,\mu})_{,\mu} ] &=  \frac{1}{2}(\Psi^{\Lambda})^2(\Delta_{r,rr}^{\Lambda}+\Delta_{\mu,\mu\mu}^{\Lambda})\nonumber\\
  +\Delta^{\Lambda}_{r}(\tilde{Z}^{\Lambda}_{,r})^2 &+\Delta^{\Lambda}_{\mu}(\tilde{Z}^{\Lambda}_{,\mu})^2,
\end{align}
where the expressions $\rho^2=r^2+a^2\mu^2$ and $\tilde{H},\tilde{V}$ don't depend on $\Lambda$. Therefore, the complex function $\tilde{Z}^{\Lambda}$ is actually expressed as
\begin{equation}
\tilde{ Z}^{\Lambda}=\tilde{\Psi}^{\Lambda}+i \tilde{\Phi}^{\Lambda}.
\end{equation}
Unlike the homogeneous case, there is $\text{Re} Z^{\Lambda}=\tilde{\Psi}^{\Lambda}$. The particularly soluton of non-homogeneous case would be
\begin{align}
  \Delta^{\Lambda}_{r}&=-(\frac{r^4}{3}+2k^2r^2), \\
  \Delta^{\Lambda}_{\mu}&=-\frac{a^2\mu^4}{3}+2k^2\mu^2,\\
  Z^{\Lambda}&=-\frac{1}{3}(r+i a\mu)^2-2k^2,
\end{align}
where $k$ is a constant and need to be determined by match condition.

Not all the particular solutions $\Delta^{\Lambda}_{r},\Delta^{\Lambda}_{\mu},Z^{\Lambda}$ plus the general homogeneous solution $\Delta^{K}_{r},\Delta^{K}_{\mu},Z^{K}$ satisfy the non-homogeneous equations. The match condition is chosen $k=a$. Thus, we obtain the final solutions
\begin{align}
\Delta_{r}&=r^2-2M r+a^2+Q_{*}^2-m^2S_{*}^2-\frac{\Lambda m^2}{3}r^2(r^2+a^2),\label{Soll1} \\
  \Delta_{\mu}&=1-\mu^2-\frac{\Lambda m^2}{3}a^2\mu^2(\mu^2-1),\label{Soll2}\\
    \tilde{\Psi}^K & =-\frac{\Delta_r-a^2\Delta_{\mu}}{r^2+a^2\mu^2},\label{SolKK1} \\
  \tilde{\Phi}^K & =-\frac{2aM\mu}{r^2+a^2\mu^2}-\frac{2 m^2\Lambda}{3}ar\mu,
\end{align}
with the same $\tilde{H}$ and $\tilde{V}$ of homogeneous solutions. The original unknown $\rho^2,\chi,q$ can also be expressed as
\begin{align}
\rho^2&=r^2+a^2\mu^2,\label{rhoSR}\\
  \chi & =\frac{(r^2+a^2\mu^2)\Delta_r^{\frac{1}{2}}\Delta_{\mu}^{\frac{1}{2}}}{(r^2+a^2)^2\Delta_{\mu}-a^2(1-\mu^2)^2\Delta_r},\label{chiSR} \\
  q &=a\frac{(r^2+a^2)\Delta_{\mu}-(1-\mu^2)\Delta_r}{(r^2+a^2)^2\Delta_{\mu}-a^2(1-\mu^2)^2\Delta_r}.\label{qSR}
\end{align}

Plugging expressions (\ref{rhoSR}) - (\ref{qSR}) into Eq. (\ref{Metr1}), one can finally obtain the metric.
If we define $\mu=-\cos\theta$, the solutions transform into the Boyer-Lindquist coordinate system, which is
\begin{align}
  ds^2=&-\frac{\rho^2\Delta_r\Delta_{\theta}}{\Sigma^2}dt^2
  +\frac{\rho^2}{\Delta_r}dr^2+\frac{\rho^2\sin^2\theta}{\Delta_{\theta}}d\theta^2\nonumber\\
  &+\frac{\Sigma^2}{\rho^2}\left(d\varphi-\frac{(r^2+a^2)\Delta_{\theta}-\sin^2\theta \Delta_{r}}{\Sigma^2}adt\right)^2,\label{BLform}
\end{align}
where
\begin{align}
  \rho^2 & =r^2+a^2\cos^2\theta, \\
  \Sigma^2 & =(r^2+a^2)^2\Delta_{\theta}-a^2\sin^4\theta\Delta_{r},\\
  \Delta_{r}&=-\frac{m^2\Lambda}{3}(r^4+a^2r^2)+r^2-2Mr+a^2+Q_{*}^2-m^2S_{*}^2,\\
  \Delta_{\theta}&=\sin^2\theta\left(1+\frac{m^2\Lambda}{3}a^2\cos^2\theta\right).
\end{align}
And the vector potentials of electromagnetic fields are
\begin{align}
    A_0&=\frac{Q_{*}r}{r^2+a^2\cos^2\theta},\\
    A_1&=-\frac{Q_{*}ra\sin^2\theta}{r^2+a^2\cos^2\theta}.
\end{align}
The reference metric $f_{ab}$ is given in appendix D.

\section{Conclusion and discussion}
In this paper, we analytical derived the rotating black hole solutions in dRGT model. The black hole family in dRGT massive gravity would contain 5 parameters: the mass $M$, the charge $Q_{*}$, the angular momentum per unit mass $a=\frac{J}{M}$, the cosmological constant term $\Lambda$ and the St\"{u}ckelberg charge $S_{*}$.  Compared to the original Kerr-Newman-Ads solution, the St\"{u}ckelberg charge $ S_* $ serves as a new characteristic, or "hair," of the black hole. While we treat the charges $ S_*$ and $Q_* $ similarly, they have entirely different physical significances. The charge $ S_* $, which is associated with the mass of the graviton, does not engage in electromagnetic interactions; it solely influences the curvature of spacetime. When a point charge $ q $ moves through this spacetime, it experiences only the gravitational force from $ S_*$ and not any electromagnetic force. Importantly, the charge $ S_* $ would still exist even in the absence of the electric charge $Q_* $. By setting \( m = 0 \), the solution (\ref{BLform}) reproduces the Kerr-Newman solution. This occurs because the field equations (\ref{EinR1}) - (\ref{EinR3}) reduce to those found in Einstein-Maxwell theory when analyzed in that limit.

When taking $a=Q_{*}=0$, we obtain the static spherical symmetric solution
\begin{align*}
  ds^2&=-fdt^2+\frac{1}{f}dr^2+r^2d\theta^2+r^2\sin^2\theta d\varphi^2,
\end{align*}
where $f=1-\frac{2M}{r}-\frac{m^2S_{*}^2}{r^2}-\frac{m^2}{3}r^2$. In Ref. \cite{Li2016-1}, we obtain the static solution with $f=1-\frac{2M}{r}-\frac{S^2}{r^{\lambda}}-\Lambda r^2$. We found that in the special case $\lambda=2$, it can be returned to the solution of this article. For other $\lambda$ values, the solution has no corresponding rotating form. Thus, the black holes in dRGT model are divided into two categories: one kind of black hole solution has spherically symmetric form only; the other solution can be extended from spherical form to axisymmetric form.

Although we can use the lemma to avoid calculating the St\"{u}ckelberg field $\phi^a$, we still cannot express $I^{ab}=g^{\alpha\beta}\partial_{\alpha}\phi^a\partial_{\beta}\phi^b$ without $\phi^a$. If someone wants to determine whether the process $I^{ab}$ involves a singularity, the lemma must be applied in reverse. In the axisymmetric case, reversing the lemma would lead to another six differential equations. Addressing these equations would be challenging. This is the primary reason we utilize the lemma to bypass the calculation of the St\"{u}ckelberg field $\phi^a$. As $ a \rightarrow 0 $, the metric (\ref{BLform}) reduces to the metric in Ref. \cite{Li2016-1}, which obtains the St\"{u}ckelberg field $ \phi^a $ and a non-singular $ I^{ab} $. Thus, the invariant $I^{ab}$ does not possess any singularities. However, in the case of $a\neq 0$, the question of whether $I^{ab}$ is singular remains an unresolved problem.

The computational process of this paper originates from Chandrasekhar \cite{Chandrasekhar:1983}, which is based on a general rotating axisymmetric metric. By utilizing symmetry, the field equations can be simplified into the Ernst equations. The process is akin to addressing spherically symmetric problems, although it involves complex mathematical techniques. Additionally, since it does not rely on excessive assumptions, this method is also applicable to modified gravity theories. Another commonly used method for obtaining axisymmetric solutions is the Newman-Janis algorithm. The solution presented in (\ref{BLform}) can also be derived using this algorithm, which is further discussed in the paper \cite{Li2025-1}.

\textbf{Acknowledgement}
This work is partially supported by the National Natural Science Foundation of China (NSFC U2031112) and the Scientific Research Foundation of Hunan University of Arts and Sciences (23BSQD237), and by Hunan Provincial Department of Education Key Laboratory of Information Detection and Intelligent Processing Technology, and by Applied Characteristic Subject of Hunan Province "Electronic Science and Technology". We also acknowledge the science research grants from the China Manned Space Project with NO. CMS-CSST-2021-A06.

\appendix
\section*{Appendix A: The derivation process 1}

Using $\Psi$ and Eqs. (\ref{qr}) and (\ref{qmu}), we express $q_{,r}$ and $q_{,\mu}$ as
\begin{align}
  -q_{,\mu} &=\frac{\Delta_{r}}{\Psi^2}[\Phi_{,r}+2\text{Im}(HH^{*}_{,r}-m^2VV^{*}_{,r})],\label{qr1}  \\
  q_{,r} & =\frac{\Delta_{\mu}}{\Psi^2}[\Phi_{,\mu}+2\text{Im}(HH^{*}_{,\mu}-m^2VV^{*}_{,\mu})].\label{qmu1}
\end{align}
The constriant Eqs. (\ref{Eq4}) and (\ref{Eq5}) can be expressed as the form
\begin{align}
  &\left(\frac{\Delta_r}{\Psi}B_{,r} \right)_{,r}+\left(\frac{\Delta_{\mu}}{\Psi}B_{,\mu} \right)_{,\mu}=\frac{\Delta_r}{\Psi^2}A_{,r}(\Phi_{,r}+2\text{Im}(HH^{*}_{,r}\nonumber\\
  &-m^2VV^{*}_{,r}))+\frac{\Delta_{\mu}}{\Psi^2}A_{,\mu}(\Phi_{,\mu}+2\text{Im}(HH^{*}_{,\mu}-m^2VV^{*}_{,\mu})),\label{A1}\\
  & \left(\frac{\Delta_r}{\Psi}A_{,r} \right)_{,r}+\left(\frac{\Delta_{\mu}}{\Psi}A_{,\mu} \right)_{,\mu}=-\frac{\Delta_r}{\Psi^2}B_{,r}(\Phi_{,r}+2\text{Im}(HH^{*}_{,r}\nonumber\\
  &-m^2VV^{*}_{,r}))-\frac{\Delta_{\mu}}{\Psi^2}B_{,\mu}(\Phi_{,\mu}+2\text{Im}(HH^{*}_{,\mu}-m^2VV^{*}_{,\mu})),\label{A2}
\end{align}
 We calculate
\begin{align}
 &\left(\frac{\Delta_{r} BB_{,r}}{\Psi} \right)_{,r}+\left(\frac{\Delta_{\mu} BB_{,\mu}}{\Psi} \right)_{,\mu}  =\frac{\Delta_{r} B_{,r}^2}{\Psi}+\frac{\Delta_{\mu} B_{,\mu}^2}{\Psi}\nonumber \\
 &+B\left(\left(\frac{\Delta_{r} B_{,r}}{\Psi} \right)_{,r}+\left(\frac{\Delta_{\mu} B_{,\mu}}{\Psi} \right)_{,\mu}\right )\nonumber \\
 & =\frac{\Delta_{r} B_{,r}^2}{\Psi}+\frac{\Delta_{\mu} B_{,\mu}^2}{\Psi}+\frac{\Delta_r}{\Psi^2}BA_{,r}(\Phi_{,r}+2\text{Im}(HH^{*}_{,r}-m^2VV^{*}_{,r}))\nonumber\\
 &+\frac{\Delta_{\mu}}{\Psi^2}BA_{,\mu}(\Phi_{,\mu}+2\text{Im}(HH^{*}_{,\mu}-m^2VV^{*}_{,\mu})).
\end{align}
Similarly, there is
\begin{align}
 &\left(\frac{\Delta_{r} AA_{,r}}{\Psi} \right)_{,r}+\left(\frac{\Delta_{\mu} AA_{,\mu}}{\Psi} \right)_{,\mu}  =\frac{\Delta_{r} A_{,r}^2}{\Psi}+\frac{\Delta_{\mu} A_{,\mu}^2}{\Psi}\nonumber\\
   &-\frac{\Delta_r}{\Psi^2}AB_{,r}(\Phi_{,r}+2\text{Im}(HH^{*}_{,r}-m^2VV^{*}_{,r}))\nonumber\\
   &- \frac{\Delta_{\mu}}{\Psi^2}AB_{,\mu}(\Phi_{,\mu}+2\text{Im}(HH^{*}_{,\mu}-m^2VV^{*}_{,\mu})).
\end{align}
Notice that $|H|^2_{,r}=2(AA_{,r}+BB_{,r})$ and $\text{Im} HH^{*}_{,r}=BA_{,r}-AB_{,r}$. Add the above two equations, one have
\begin{align}
  &\Psi[(\Delta_r|H|^2_{,r})_{,r}+ (\Delta_{\mu}|H|^2_{,\mu})_{,\mu}] = 2\Psi(\Delta_r|H_{,r}|^2+\Delta_{,\mu}|H_{,\mu}|^2) \nonumber\\
   &+ \Delta_r\{\Psi_{,r}|H|^2_{,r}+2\text{Im}HH^{*}_{,r}(\Phi_{,r}+2\text{Im}(HH^{*}_{,r}-m^2VV^{*}_{,r}))   \} \nonumber\\
   &+\Delta_{\mu}\{\Psi_{,\mu}|H|^2_{,\mu}+2\text{Im}HH^{*}_{,\mu}(\Phi_{,\mu}+2\text{Im}(HH^{*}_{,\mu}-m^2VV^{*}_{,\mu}))   \}. \label{I}
\end{align}
One can also calculate
\begin{align}
  \left(\frac{\Delta_r(AB_{,r}-BA_{,r})}{\Psi} \right)_{,r} &=A\left(\frac{\Delta_rB_{,r}}{\Psi} \right)_{,r}-B\left(\frac{\Delta_rA_{,r}}{\Psi} \right)_{,r}.
\end{align}
By direct calculation and Using Eqs. (\ref{A1}) and (\ref{A2}), we obtain
\begin{align}
 &2\Psi[ (\Delta_r\text{Im} HH^{*}_{,r})_{,r}+ (\Delta_{\mu}\text{Im} HH^{*}_{,\mu})_{,\mu} ]= - \Delta_r|H|^2_{,r}(\Phi_{,r}\nonumber  \\
   &+2\text{Im}(HH^{*}_{,r}-m^2VV^{*}_{,r})) -\Delta_{\mu}|H|^2_{,\mu}(\Phi_{,\mu}+2\text{Im}(HH^{*}_{,\mu}\nonumber  \\
   &-m^2VV^{*}_{,\mu}))+2\Delta_r\Psi_r \text{Im} HH^{*}_{,r}+ 2\Delta_{\mu}\Psi_{\mu} \text{Im} HH^{*}_{,\mu}.\label{II}
\end{align}

For $V$, we also have
\begin{align}
  &\Psi[(\Delta_r|V|^2_{,r})_{,r}+ (\Delta_{\mu}|V|^2_{,\mu})_{,\mu}] = 2\Psi(\Delta_r|V_{,r}|^2+\Delta_{,\mu}|V_{,\mu}|^2) \nonumber\\
   &+ \Delta_r\{\Psi_{,r}|V|^2_{,r}+2\text{Im}VV^{*}_{,r}(\Phi_{,r}+2\text{Im}(HH^{*}_{,r}-m^2VV^{*}_{,r}))   \} \nonumber\\
   &+\Delta_{\mu}\{\Psi_{,\mu}|V|^2_{,\mu}+2\text{Im}VV^{*}_{,\mu}(\Phi_{,\mu}+2\text{Im}(HH^{*}_{,\mu}-m^2VV^{*}_{,\mu}))   \}, \label{III}\\
   &2\Psi[ (\Delta_r\text{Im} VV^{*}_{,r})_{,r}+ (\Delta_{\mu}\text{Im} VV^{*}_{,\mu})_{,\mu} ]= -\Delta_r|V|^2_{,r}(\Phi_{,r}\nonumber  \\
   &+2\text{Im}(HH^{*}_{,r}-m^2VV^{*}_{,r}))-\Delta_{\mu}|V|^2_{,\mu}(\Phi_{,\mu}+2\text{Im}(HH^{*}_{,\mu}\nonumber  \\
   & -m^2VV^{*}_{,\mu}))+2\Delta_r\Psi_r \text{Im} HH^{*}_{,r}+ 2\Delta_{\mu}\Psi_{\mu} \text{Im} HH^{*}_{,\mu}.\label{IV}
\end{align}

\section*{Appendix B: The derivation process 2}  

The equation (\ref{equation1}) can be transform to
\begin{align}
  &\Psi\{[\Delta_r(\Phi_{,r}+2\text{Im}(HH^{*}_{,r}-VV^{*}_{,r}))]_{,r}+ [\Delta_{\mu}(\Phi_{,\mu}+2\text{Im}(HH^{*}_{,\mu}\nonumber\\&-VV^{*}_{,\mu}))]_{,\mu}\}=
  2\Delta_r\Psi_{,r}(\Phi_{,r}+2\text{Im}(HH^{*}_{,r}-VV^{*}_{,r}))\nonumber\\&+2\Delta_{\mu}\Psi_{,\mu}(\Phi_{,\mu}+2\text{Im}(HH^{*}_{,\mu}-VV^{*}_{,\mu})).
\end{align}
Pluging (\ref{II}) and (\ref{IV}) into above equation, one can derive
\begin{align}
  &\Psi[(\Delta_{r}\Phi_{,r} )_{,r}+ (\Delta_{\mu}\Phi_{,\mu})_{,\mu}] =\Delta_r\{\Phi_{,r}(2\Psi+|H|^2-m^2|V|^2)_{,r}\nonumber \\
  & +2(\Psi+|H|^2-m^2|V|^2)_{,r}\text{Im}(HH^{*}_{,r}-VV^{*}_{,r}) \}\nonumber \\
  & +\Delta_{\mu}\{\Phi_{,\mu}(2\Psi+|H|^2-m^2|V|^2)_{,\mu} \nonumber \\
  &+2(\Psi+|H|^2-m^2|V|^2)_{,\mu}\text{Im}(HH^{*}_{,\mu}-VV^{*}_{,\mu}) \}.\label{V}
\end{align}

\section*{Appendix C: The derivation process 3} 
We express
\begin{equation}
  \ln \chi=\frac{1}{2}(\ln \Delta_{r}+\ln \Delta_{\mu})-\ln\Psi.
\end{equation}
Then, one has
\begin{equation}
  (\Delta_r (\ln\chi)_{,r})_{,r}=\frac{1}{2}\Delta_{r,rr}-\left(\Delta_r\frac{\Psi_{,r}}{\Psi}\right)_{,r}.
\end{equation}
The Eq. (\ref{CEq1}) can be transform to
\begin{align}
  &\frac{1}{2}(\Delta_{r,rr}+\Delta_{\mu,\mu\mu})-\left(\Delta_r\frac{\Psi_{,r}}{\Psi}\right)_{,r}-\left(\Delta_{\mu}\frac{\Psi_{,\mu}}{\Psi}\right)_{,\mu}
  -\frac{\Psi^2}{\Delta_{\mu}}q_{,r}^2\nonumber\\
  &-\frac{\Psi^2}{\Delta_{r}}q_{,\mu}^2=\frac{2}{\Psi}[\Delta_r(|H_{,r}|^2-m^2|V_{,r}|^2)+ \Delta_{\mu}(|H_{,\mu}|^2-m^2|V_{,\mu}|^2)].
\end{align}
Pluging Eqs. (\ref{qr1}) and (\ref{qmu1}), one can obtain
\begin{align}
  &\Psi[(\Delta_{r}\Psi_{,r} )_{,r}+ (\Delta_{\mu}\Psi_{,\mu})_{,\mu}] =\frac{\Psi^2}{2}(\Delta_{r,rr}+\Delta_{\mu,\mu\mu})+\Delta_r \Psi_{,r}^2\nonumber \\
   &+\Delta_{\mu}\Psi_{,\mu}^2-2\Psi[\Delta_r(|H_{,r}|^2-m^2|V_{,r}|^2)\nonumber \\
   &+\Delta_{\mu}(|H_{,\mu}|^2-m^2|V_{,\mu}|^2)] -\Delta_r[ \Phi_{,r}+2\text{Im}(HH^{*}_{,r}-VV^{*}_{,r})  ]^2\nonumber \\
   &-\Delta_{\mu}[ \Phi_{,\mu}+2\text{Im}(HH^{*}_{,\mu}-VV^{*}_{,\mu}) ]^2.\label{VI}
\end{align}

\section*{Appendix D: The derivation of the reference metric $f_{ab}$} 

Functions $A',B'$ are given by the expressions $A'=-S_{*}\frac{ra(1-\mu^2)}{\rho^2}$,  $B'=-S_{*}\frac{\mu(r^2+a^2)}{\rho^2}$ in solutions (\ref{BLform}), similar to the expressions $A=-Q_{*}\frac{ra(1-\mu^2)}{\rho^2},B=-Q_{*}\frac{\mu(r^2+a^2)}{\rho^2}$. Plugging them into Eqs. (\ref{TmS1}) - (\ref{TmS2}), we obtain
\begin{align}
&T^{(\mathcal{K})}_S{}^0{}_0=-T^{(\mathcal{K})}_S{}^1{}_1=\nonumber\\
&-S_{*}^2e^{-2(\psi+\mu_2+\mu_3)}\frac{e^{2\mu_2}(r^2+a^2)^2+a^2e^{2\mu_3}(\mu^2-1)^2}{\rho^4},\label{TmSsol1}\\
&T^{(\mathcal{K})}_S{}^2{}_2=-T^{(\mathcal{K})}_S{}^3{}_3=\nonumber\\&-S_{*}^2e^{-2(\psi+\mu_2+\mu_3)}\frac{e^{2\mu_2}(r^2+a^2)^2-a^2e^{2\mu_3}(\mu^2-1)^2}{\rho^4},\\
&T^{(\mathcal{K})}_S{}^0{}_1=2aS_{*}^2e^{-(2\psi+\mu_2+\mu_3)}\frac{(\mu^2-1)(r^2+a^2)}{\rho^4},\\
&T^{(\mathcal{K})}_S{}^2{}_3=0,\label{TmSsol2}
\end{align}
Notice that $T^{(\mathcal{K})}{}^{a}{}_{b}=T^{(\mathcal{K})}_S{}^{a}{}_{b}+T^{(\mathcal{K})}_{\Lambda}{}^{a}{}_{b}=T^{(\mathcal{K})}_S{}^{a}{}_{b}+\Lambda \delta^a_b$. Therefore, we got the expressions of $T^{(\mathcal{K})}{}^{a}{}_{b}$ given by solutions (\ref{BLform}). At the same time, the expressions of $T^{(\mathcal{K})}{}^{a}{}_{b}$ are also given by the reference metric $f_{ab}$ (\ref{TK1}) - (\ref{TK2}). Let them equal, we obtain six algebraic equations. Mathematically, the six equations will uniquely determine the six unknowns $f_{00},f_{01},f_{11},f_{22},f_{23},f_{33}$. However, due to the appearance of the fourth power, the analytical expressions $f_{00},f_{01},f_{11},f_{22},f_{23},f_{33}$ are difficult to obtain. This problem further makes it difficult to analyze the singularity of $I^{ab}$.

In the special case that $c_3=c_4=0$, it is possible to analytical express the reference metric $f_{ab}$. A more concise derivation is as follows. Defining
\begin{align*}
  K_1 & =S_{*}e^{-\psi-\mu_3}\frac{r^2+a^2}{\rho^2}, \\
  K_2 & =S_{*}ae^{-\psi-\mu_2}\frac{1-\mu^2}{\rho^2},
\end{align*}
We can reexpress
\begin{align*}
T^{(\mathcal{K})}{}^0{}_0&=-(K_1^2+K_2^2)+\Lambda,\\
T^{(\mathcal{K})}{}^1{}_1&=(K_1^2+K_2^2)+\Lambda,\\
T^{(\mathcal{K})}{}^2{}_2&=-(K_1^2-K_2^2)+\Lambda,\\
T^{(\mathcal{K})}{}^3{}_3&=(K_1^2-K_2^2)+\Lambda,\\
T^{(\mathcal{K})}{}^0{}_1&=-T^{(\mathcal{K})}{}^1{}_0=-2K_1K_2.
\end{align*}
Inserting
\begin{equation}
  \gamma=\left(
           \begin{array}{cccc}
             \gamma_{00} & \gamma_{01} & 0 & 0 \\
             \gamma_{10} & \gamma_{11} & 0 & 0\\
             0 & 0 & \gamma_{22} & 0 \\
             0 & 0 & 0& \gamma_{33} \\
           \end{array}
         \right),
\end{equation}
into Eqs. (\ref{TK}), one obtain
\begin{align*}
T^{(\mathcal{K})}{}^0{}_0&=-\frac{3}{2} - q \gamma_{01} + \gamma_{11} + \gamma_{22} + \gamma_{33},\\
T^{(\mathcal{K})}{}^1{}_1&=-\frac{3}{2} + \gamma_{00} + q \gamma_{01} +\gamma_{22} + \gamma_{33},\\
T^{(\mathcal{K})}{}^2{}_2&=-\frac{3}{2} + \gamma_{00} + \gamma_{11} + \gamma_{33},\\
T^{(\mathcal{K})}{}^3{}_3&=-\frac{3}{2} + \gamma_{00} + \gamma_{11} + \gamma_{22},\\
T^{(\mathcal{K})}{}^0{}_1&=-e^{\nu - \psi} \gamma_{01},\\
T^{(\mathcal{K})}{}^1{}_0&=e^{-\nu + \psi} (-\gamma_{10} +
   q (\gamma_{00} + q \gamma_{01} - \gamma_{11})).
\end{align*}
Letting them equal, one obtains
\begin{align}
  \gamma_{00} &=\frac{1}{2} + K_1^2 + K_2^2 - 2 e^{-\nu + \psi} K_1 K_2 q + \frac{\Lambda}{3}, \\
  \gamma_{11} & =\frac{1}{2} - K_1^2 - K_2^2 + 2 e^{-\nu + \psi} K_1 K_2 q + \frac{\Lambda}{3},\\
  \gamma_{22} & =\frac{1}{2} + K_1^2 - K_2^2 + \frac{\Lambda}{3},\\
  \gamma_{33} & =\frac{1}{2} - K_1^2 + K_2^2 + \frac{\Lambda}{3},\\
  \gamma_{01} & =2 e^{-\nu + \psi} K_1 K_2,\\
   \gamma_{10} &=-2 e^{\nu - \psi} K_1 K_2 + 2 (K_1^2 + K_2^2) q -
 2 e^{-\nu + \psi} K_1 K_2 q^2 .
\end{align}
Within Eq. (\ref{gamma}) and $\phi^a=x^{\mu}\delta_{\mu}^a$, we have $f_{\alpha\beta}=g_{\alpha\mu}\gamma^2{}^{\mu}{}_{\beta}$. The directly calculation shows
\begin{align}
  f_{00} &=-\frac{1}{36} e^{2 \nu} (3 + 6 (K_1 - K_2)^2 + 2 \Lambda) (3 +
     6 (K_1 + K_2)^2 \nonumber\\
   &+ 2 \Lambda) +
\frac{2}{3} e^{\nu -
   3 \psi} (1 + e^{4 \psi}) K_1 K_2 (3 + 2 \Lambda) q \nonumber\\
   &+
\frac{1}{36} (-24 e^{-2 \psi} (K_1^2 + K_2^2) (3 + 2 \Lambda) +
    e^{2 \psi} (3 \nonumber\\
       &+ 6 (K_1 - K_2)^2 + 2 \Lambda) (3 +
       6 (K_1 + K_2)^2 + 2 \Lambda)) q^2 \nonumber\\
   &-
\frac{2}{3} (e^{-\nu - \psi} (-1 + e^{4 \psi}) K_1 K_2 (3 +
      2 \Lambda)) q^3  \\
  f_{11} & =\frac{1}{36} e^{2 \psi} (-3 + 6 (K_1 - K_2)^2 - 2 \Lambda) (-3 +
    6 (K_1 + K_2)^2\nonumber\\ & - 2 \Lambda) +
\frac{2}{3} e^{-\nu - \psi} (-1 + e^{4 \psi}) K_1 K_2 (3 + 2 \Lambda) q,\\
f_{22}&=e^{2 \mu_2} (\frac{1}{2} + K_1^2 - K_2^2 + \frac{\Lambda}{3} )^2,\\
f_{33}&=e^{2 \mu_3} (\frac{1}{2} - K_1^2 + K_2^2 + \frac{\Lambda}{3} )^2,\\
f_{01}&=-\frac{2}{3} (e^{\nu + \psi} K_1 K_2 (3 + 2 \Lambda)) -
 \frac{1}{36} (e^{-2 \psi} (-3 \nonumber\\ &+ 6 (K_1 - K_2)^2 - 2 \Lambda) (-3 +
      6 (K_1 + K_2)^2 - 2 \Lambda)) q \nonumber\\ &+
\frac{2}{3} e^{-\nu - \psi} (-1 + e^{4 \psi}) K_1 K_2 (3 +
    2 \Lambda) q^2.
\end{align}

\end{document}